# NUMERICAL AND ANALYTICAL STUDIES ON MODEL GRAVITATING SYSTEMS

By

KIM A. GARGAR

A Masteral Thesis Submitted to the
National Institute of Physics
College of Science
University of the Philippines
Diliman, Quezon City

In Partial Fulfillment of the Requirements
for the Degree of
Master of Science in Physics

OCTOBER 2002

# ABSTRACT

In this thesis we study the evolution of systems of concentric shells interacting gravitationally and in the process (1) propose and implement a nearly energy-conserving numerical integration scheme for evolving the concentric spherical shells systems with 1024 particles or less; (2) look at the possibility of chaos in few shell systems; and (3) study the evolution of many shell systems in the Vlasov limit.

The proposed numerical integration scheme is a nearly energy conserving hybrid of the Verlet and modified Euler-Cromer integration schemes. The superior performance of the new scheme is exhibited through total energy versus time plots from the new scheme and the following integrators: Euler, Euler-Cromer, modified Euler-Cromer, Verlet, fourth-order Runge-Kutta. The energy conservation requirement is critical for long-time simulations.

The rotational 2-shell spherical system is investigated in detail using the hybrid numerical integration scheme. Plots of time-series, phase space projections, Poincare sections, power spectra, and Lyapunov exponents are obtained for the system. These diagnostic tools, taken together, clearly show the chaotic nature of the rotational 2-shell system. Three types of periodic orbits are observed: collapsed, one-point, and three-point periodic orbits. We believe that the three-point periodic orbits result from a rotation-induced bifurcation. Four types of quasiperiodic orbits are also observed. Three of these are a result of slight changes in the initial conditions corresponding to the three types of periodic orbits. The fourth type of quasiperiodic orbit separates the chaotic region from the non-chaotic regions in phase space.

The short-time evolution of collisionless spherical shells system is studied using both numerical and analytical methods. For the short time evolution of *N*-shells system ($N$ = 32, 64, 128, 256, 512, 1024) the results of the Verlet and the hybrid Verlet-modified Euler-Cromer schemes are identical if the evolution time is so short that there are no collisions. Number density plots are produced from the numerical simulations. Approximate expressions for the short-time evolution of the collisionless rotational shells system are obtained using Vlasov-Poisson perturbation theory in the high-virial limit. The agreement between the analytical results and numerical results for finite shells systems improves as the number of shells in the system increases.

# Table of Contents

# List of cited Figures

**Figure 1**. Total energy of a 2-shell system evolving under 7 different numerical integration methods.
**Figure 2**. Enlarged view of the result (dotted line) shown in Figure 1.
**Figure 3**. Longer time fluctuations in the total energy of a 2-shell system with $\delta t = 0.001$.
**Figure 4**. Fluctuations in the total energy of a 2-shell system for different time-step size, $\delta t$.
**Figure 5**. Two-sided Poincare surface of section of a 2-shell system.
**Figure 6**. Projection at $v_1v_2$-plane of two-sided Poincare sections showing aperiodic solutions.
**Figure 7**. Power spectrum of $v_1$ corresponding to the chaotic region in Figure 5 consisting of widely scattered points.
**Figure 8**. Separation distance as a function of time between two chaotic trajectories of a 2-shell system showing the sensitivity to initial condition.
**Figure 9**. Periodic solution of a 2-shell system projected at $r_1v_1$-plane.
**Figure 10**. The three-point periodic solution projected at $r_1v_1$-plane.
**Figure 11**. Initial phase-space distribution $f(r, v, 0) = n(r)\varphi(v)$.
**Figure 12**. Plot of number density given by equation 4.34.
**Figure 13**. Number density at $t = T = z/s$ of $N$-shell systems initially ($t = 0$) described by equations 4.30 and 4.31 and with virial ratio 20.

# Chapter 1
# Introduction

## 1.1 Motivation

In this thesis, model self-gravitating systems involving concentric spherical shells are investigated.

There are several motivations for studying model gravitating systems:

First, it is commonly believed that the present observed structures in the universe such as galaxies and star clusters are a result of the evolution starting from initial density fluctuations in the primordial universe [1-5]. Simulating the evolution of these structures is difficult because the usual complications involving nonlinearity and many-particle dynamics are compounded by other issues like gravitational singularities and the possibility of escape from the system. One can always evolve simple model systems for much longer times than is practicable for the more complex three-dimensional systems. The hope is that new phenomena we observe in the model systems may alert us on the possible existence of similar phenomena for the more realistic systems.

Second, these model systems can be, and have been, used to model astrophysical systems. For instance, the parallel sheets model has been used to model the behavior of halo stars [6], while the concentric spherical shells model has been used to model globular clusters [7].

Third, few body systems involving these models have been observed to exhibit chaos [8-13].

Fourth, because of their simplicity it is possible to cross-validate numerical *N*-body and continuum statistical mechanics approaches using these model systems.

The rest of the chapter is organized as follows. In Section 1.2, we review previous work on concentric spherical shells systems. In Section 1.3, we note several problems left unaddressed in the concentric spherical shells literature and state the problems addressed in this thesis.

## 1.2 The concentric shells model

The concentric shells system models a collection of masses that are distributed in a spherically symmetric manner in space. Every shell in the system is composed of point masses -- all moving with identical radial velocity components. If the tangential component of the velocity of the point masses is identically zero, we have a non-rotational or irrotational spherical shells system. If the magnitude of the tangential component of the velocity of the points on a spherical shell is non-zero but still identical, we have a rotational spherical shells system. In the astronomical context, the point masses may represent stars.

The concentric shells system was introduced by Campbell [14-15] and used later by Henon [16-17] as a model of a spherically symmetric star cluster. Bouvier and Janin [18] investigated the evolution of a self-gravitating spherical system during its preliminary stage of orbital mixing. Subsequent studies on the concentric shells system have dealt with equilibrium properties, approach to equilibrium, nonlinear dynamics, their role as model for globular clusters, the effects of rotation, etc.

Because the irrotational spherical shells system is simpler than its rotational counterpart, it is not surprising that it has been studied in greater detail. Various aspects of its equilibrium and steady state behavior have been studied. Henon [19] obtained spherical steady states and investigated numerically the stability of these states with respect to spherical disturbances. Recently, the equilibrium distribution of the concentric shells system was derived using a mean-field theoretic approach [20]. Gravitational phase transitions have also been studied both analytically and numerically [21].

Once the steady state or equilibrium behavior is known it is natural to ask at what rate the system approaches equilibrium starting from a non-equilibrium state. It has been found [23] that the approach to equilibrium is much faster in the concentric shells system than in the parallel sheets system [24-36].

This may be because the region of phase space exhibiting chaos is larger for the spherical shells system compared to the counterpart parallel sheets system [8,11]. Miller and Youngkins [13] have observed chaos in an irrotational concentric shells system with only two shells.

The concentric shells system has also been used to model globular clusters. Fanelli and Ruffo [7] used an irrotational shells system as a model for studying the evaporation process in globular clusters. They, however, had to introduce an inner reflecting barrier to shield the system from a singularity at the center. They were able to show that the radius of the inner barrier had an effect on the evaporation process.

The inherent singularity of the irrotational concentric spheres system complicates both analytical and numerical calculations. One possible way to deal with the singularity -- the introduction of an inner barrier -- has already been mentioned. This procedure, however, may be criticized for being 'artificial'. A second way to deal with the singularity is by allowing the masses composing the shells to have velocity components in the tangential direction. This rotational component effectively shields the singularity.

The effects of rotation in spherical shells systems have only been studied recently. Klinko and Miller [37-39] studied rotation-induced phase transitions in spherical shells system by calculating and analyzing the equilibrium distribution of the shells system in the continuum limit. The existence of chaos in the rotational 2-shells system was asserted by a Russian group [40] but conclusive evidence remains to be shown since they only relied on qualitative observation of the time-series plots.

## 1.3 Open problems

In reviewing the literature, we noted several unaddressed problems.

*Is the rotational 2-shell system chaotic?*

Miller and Youngkins [13] demonstrated that a system made up of two *irrotational* concentric spherical shells restricted to a region bounded by an inner and an outer spherical barrier exhibits chaos. Meanwhile, Barkov, Belinski, and Bisnovatyi-Kogan [40] asserted that a *rotational* two-shells system, with no artificial barriers, exhibits chaos. This assertion was made on the basis of time-series plots of the shells' positions. However, the visual appearance of time-series plots alone is insufficient to prove such assertion -- for example the time series plot of a quasiperiodic system may appear deceptively similar to that of a chaotic system. A more detailed analysis of the rotational 2-spheres system is needed to ascertain the existence of chaos and to characterize its phase space structure. A natural consequence of this analysis is the identification of regions of phase space corresponding to fixed point, periodic, and quasiperiodic behavior in addition to the chaotic regions we are mainly after.

*Will analytical calculations for the continuum limit be consistent with N-body simulations of the rotational spherical shells system?*

The equilibrium and non-equilibrium behavior of irrotational shells systems have been studied using numerical and analytical methods. The equilibrium behavior of rotational spherical systems has been studied using numerical and analytical methods, while its non-equilibrium behavior has been studied using numerical methods only. Equally important analytical studies of rotational spherical shells systems in the continuum limit are lacking. The availability of analytical results in the continuum limit will help validate $N$-body simulations of the rotational spherical shells systems.

The above problems will be addressed in this thesis in the following way.

First, a method for numerically evolving rotational spherical shells systems is needed to address both problems. The method should not yield unphysical results like non-conservation of energy. At the same time the method should be fast. In Chapter 2, we propose a new numerical integration scheme that satisfies both requirements. We compare computed total energy obtained from traditional numerical integrators with those obtained with the new numerical integrator and demonstrate the superiority of the new method over the traditional methods. We would then use this method in our subsequent simulations. In particular, we will employ the new method to simulate the rotational N-shell system for N =2, 32, 64, 128, 256, 512, and 1024 shells. The N=2 simulations are used in Chapter 3, while the higher N simulations are used in Chapter 4.

Second, to ascertain chaos in rotational two-shells systems we need to provide more concrete evidence of chaos beyond mere visual inspection of time-series plots. We do this in Chapter 3 by utilizing several standard tools in analyzing the dynamics of nonlinear systems. We use time series obtained using the numerical integration scheme developed in Chapter 2 for $N = 2$ under various initial conditions. By constructing the Poincare section of different orbits, we can determine the phase-space structure of the rotational 2-shell system from which different types of solutions (that is, periodic, quasiperiodic and chaotic) are revealed. Further evidence of chaos is exhibited by the sensitivity to initial conditions and by computing the power spectrum of the time-series.

Third, before we can compare analytical calculations with $N$-body simulations, we have to perform the analytical calculations and run the numerical simulations. The analytical calculations start with the Vlasov-Poisson or collisionless Boltzmann equation and proceed by implementing a perturbation method to obtain an approximate solution to the equation. The ultimate result is a first-order approximation of the number density of the shells system during the early collisionless stage of its evolution. The numerical simulations employ the numerical integrator developed in Chapter 2. To verify our analytical approximations, we evolve the $N$-shell system for times short enough for collisions not to occur and then construct the number density to be compared with the analytical result. The numerical simulations are done for $N = 32$, 64, 128, 256, 512, and 1024. These are all reported in Chapter 4.

We conclude the thesis in Chapter 5 by summarizing the new results and mentioning possible future work.

# Chapter 2
# Numerical techniques for one-dimensional *N*-body simulations

Our goals in this chapter are to present numerical techniques which may be used in simulating one-dimensional *N*-particle systems and to assess their performance in simulating the shells system. The numerical technique which best satisfies the accuracy and speed requirements will be used to study various concentric shells system in the succeeding chapters.

Numerical simulation of classical one-dimensional many-body systems can be done using either event-driven algorithms or numerical integration routines. In Section 2.1, we describe an event-driven algorithm for the concentric shells systems and discuss the difficulties involved in implementing it. In Section 2.2, we compare the performance of traditional numerical integration schemes with a proposed hybrid numerical integrator for simulating the rotational 2-shell system. The superiority of the new scheme is established.

## 2.1 The event-driven algorithm

The event-driven algorithm works as follows: The phase-space coordinates (i.e. position and velocity) of every particle at some given time $t_0$ are known. The *crossing times* or the time required for adjacent pairs of particles to collide or cross is computed. The system is then evolved up to the minimum of these crossing times.

For some one-dimensional systems, the crossing times depend only on the initial phase-space coordinates. For example, in the parallel sheets system, the crossing times depend only on the difference between the two particles' velocities, positions and accelerations [10, and references therein].

Furthermore, the acceleration of a sheet does not depend continuously on positions and experience discrete jumps only whenever the particle (sheet) crosses with another particle. For the shells system, a particle's (shell's) acceleration depends only on its position (radius) and on the number of shells with smaller radius. This makes it possible to calculate the crossing time in terms of the initial and final position.

### 2.1.1 Total energy of the rotating shell model

The rotating shell model consists of *N* concentric spherical shells. The total angular momentum of the *i*th shell is zero but the sum of the magnitude of the angular momenta of the infinitesimal masses making up the shell is $L_i$. The infinitesimal masses making up each shell all have identical angular speeds. Note that a single rotating spherical shell is acted upon by its *self-gravitation* plus the gravitational force due to the other shells. The energy of the *i*th shell is thus equal to

$$E_i = \frac{1}{2} m_i v_i^2 + \frac{L_i^2}{2m_i r_i^2} - \frac{1}{2}\frac{Gm_i^2}{r_i} - n_i \frac{Gm_i^2}{r_i} \tag{2.1}$$

where the first term is the kinetic energy due to its radial motion (i.e. $v_i$ is the radial velocity of the *i*th shell), the second term is the rotational kinetic energy (i.e. $L_i^2$ is the square of the angular momentum of the shell, the third term is due to self-gravitation, and the last term is the interaction potential energy which effectively is caused only by the shells inside the *i*th shell (i.e. $n_i$ is the number of shells inside the *i*th shell). This gives a total energy

$$E = \sum_{i=1}^{N} \left\{ \frac{1}{2} m_i v_i^2 + \frac{L_i^2}{2m_i r_i^2} - \frac{Gm_i}{r_i} \left[ \frac{1}{2} + \sum_{j=1}^{N} H\left(r_i - r_j\right) \right] \right\} \tag{2.2}$$

where $H(r)$ is the Heaviside function defined as follows:

$$H(r) = \begin{cases} 1 & r > 0 \\ 0 & r < 0 \\ (\text{undefined for } r = 0) \end{cases}$$

### 2.1.2 The algorithm

The event-driven algorithm relies on the existence of a method to explicitly calculate the crossing times between two particles in an *N*-body system. In the one-dimensional case, we only need to consider adjacent particles in between succeeding crossing events. This makes the method ideal for systems where the crossing time is expressible in terms of the given phase-space coordinates of the two particles before and during crossing like that in parallel sheets system and ideal gas.

This is also true for the shells system in that we can write explicitly the crossing time in terms of the phase-space coordinates. However, unlike in the sheets system, the process is not straightforward. We need to obtain first the value of the crossing radius before the crossing time is calculated. This requirement makes the entire event-driven algorithm slow and makes it prone to truncation errors during crossing radius calculations. For this reason we do not pursue further the event driven simulations for the rotational shells system.

## 2.2 Numerical integration algorithms for few shell systems

Traditional numerical integration methods may be used for simulating the dynamics of a given system. These numerical methods approximate the continuous behavior of a dynamical system by replacing the differential equations with finite difference expressions. Examples of these methods are the Euler, Verlet, Runge-Kutta, Euler-Croemer, and modified Euler-Croemer algorithms (See Appendix A). Unfortunately, these traditional numerical integrators exhibit unphysical behavior such as non-conservation of energy even though the simulated system is known to be conservative [41].

In this section we present the results of traditional numerical integrators and those of a hybrid numerical integrator that is nearly energy conserving. The new integrator is a hybrid of the Verlet and the modified Euler-Cromer algorithms. In the Verlet method the computed total energy increases after almost every crossing. In the modified Euler-Cromer method the computed total energy decreases after almost every crossing. The opposite behaviors exhibited by the Verlet and modified Euler-Cromer methods after shell crossings can be exploited to yield a hybrid algorithm that is superior to both integrators in terms of the energy conservation requirement. The hybrid algorithm works as follows: whenever the total energy is greater than or equal to the real total energy we use the modified Euler-Cromer routine. Otherwise, we use the Verlet routine. The proposed hybrid algorithm can be expressed in the following pseudocode:

```
if (computed total energy >= true total energy)
     modified Euler-Cromer routine
else % computed total energy < true total energy
     Verlet routine
```

We now look at the time evolution of the computed total energy of the 2-shell system for several numerical integrators. Figure 1 shows the energy of a 2-shell system[1] using seven different numerical integration routines. Observe that the fourth-order Runge-Kutta routine is not exempted from the finite step size effect during crossings -the finite step size leads to an increase in the computed total energy. The Euler-Cromer method is more stable than the other traditional numerical integrators but is less stable than the hybrid algorithm which did not monotonously increase the computed total energy of the 2-shell system.

[1] The dynamics is that of the 1-point periodic orbit as discussed in Chapter 3.

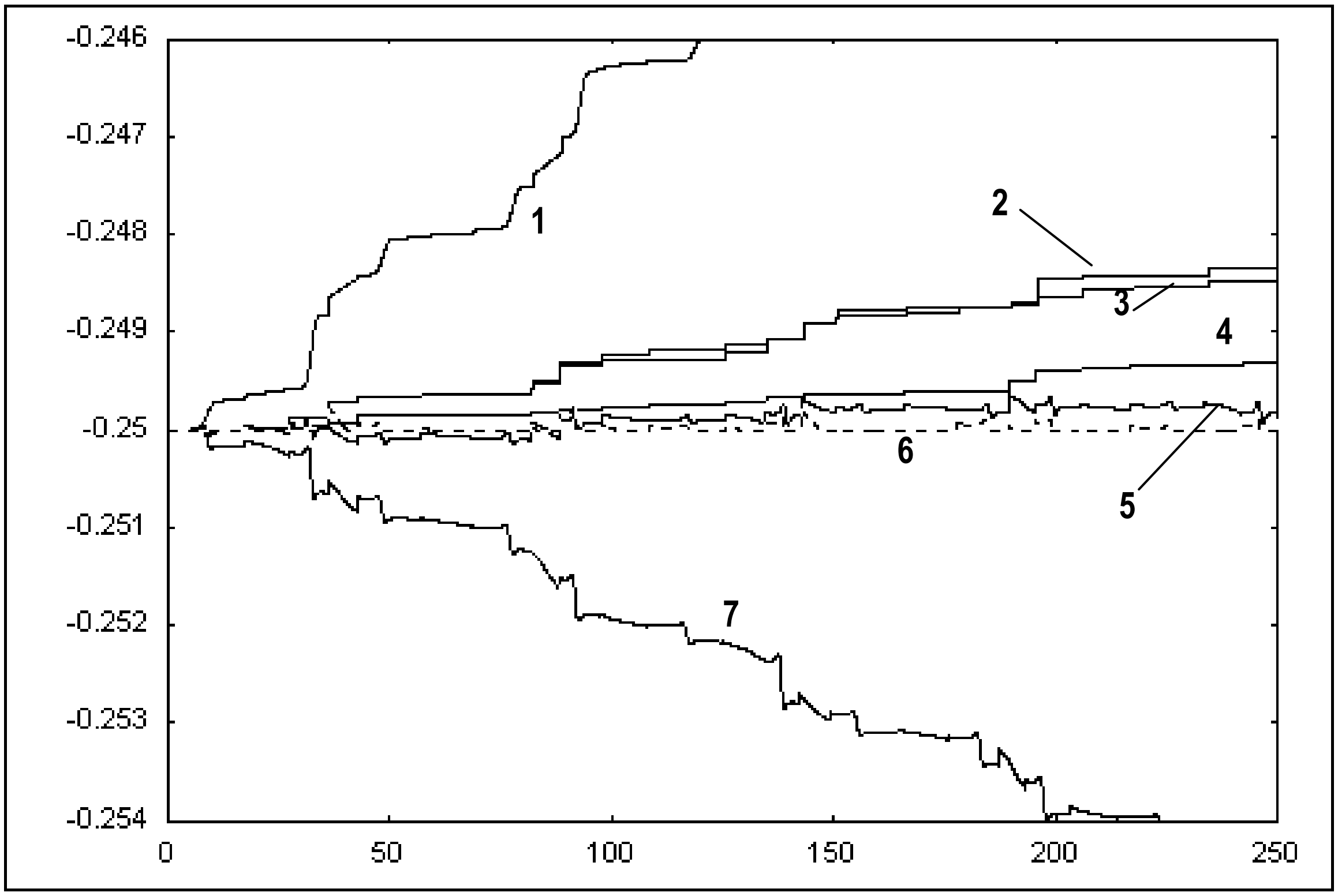

**Figure 1**. Total energy of a 2-shell system evolving under 7 different numerical integration methods, namely: 1 – Euler, 2 – Verlet, 3 – $4^{th}$-order Runge-Kutta, 4 – $4^{th}$-order Runge-Kutta (half step size), 5 – Euler-Cromer, 6 – (dotted line) Hybrid of 2 and 7, and 7 – Modified Euler-Cromer.

Having clearly shown the superiority of the new algorithm, we now explore it further. Using different time steps ($\delta t$ = 0.1, 0.05, 0.01, 0.005, 0.001), we simulate the same system using the hybrid algorithm. The result is summarized in Figure 4 and Table 1.

In this particular example, the running time behaves approximately as $C/\delta t$ while the fluctuation from the real total energy scales linearly with $\delta t$. More importantly, the total energy of the system does not deviate permanently from the true value except for fluctuations as evident in Figure 3. This cannot be achieved in ordinary methods like the ones mentioned above no matter how small the step size is.

In the next chapter, Chapter 3, we use the nearly energy-conserving algorithm in simulating the dynamics of a 2-shell system. The results of these simulations will be used as input for further analysis.

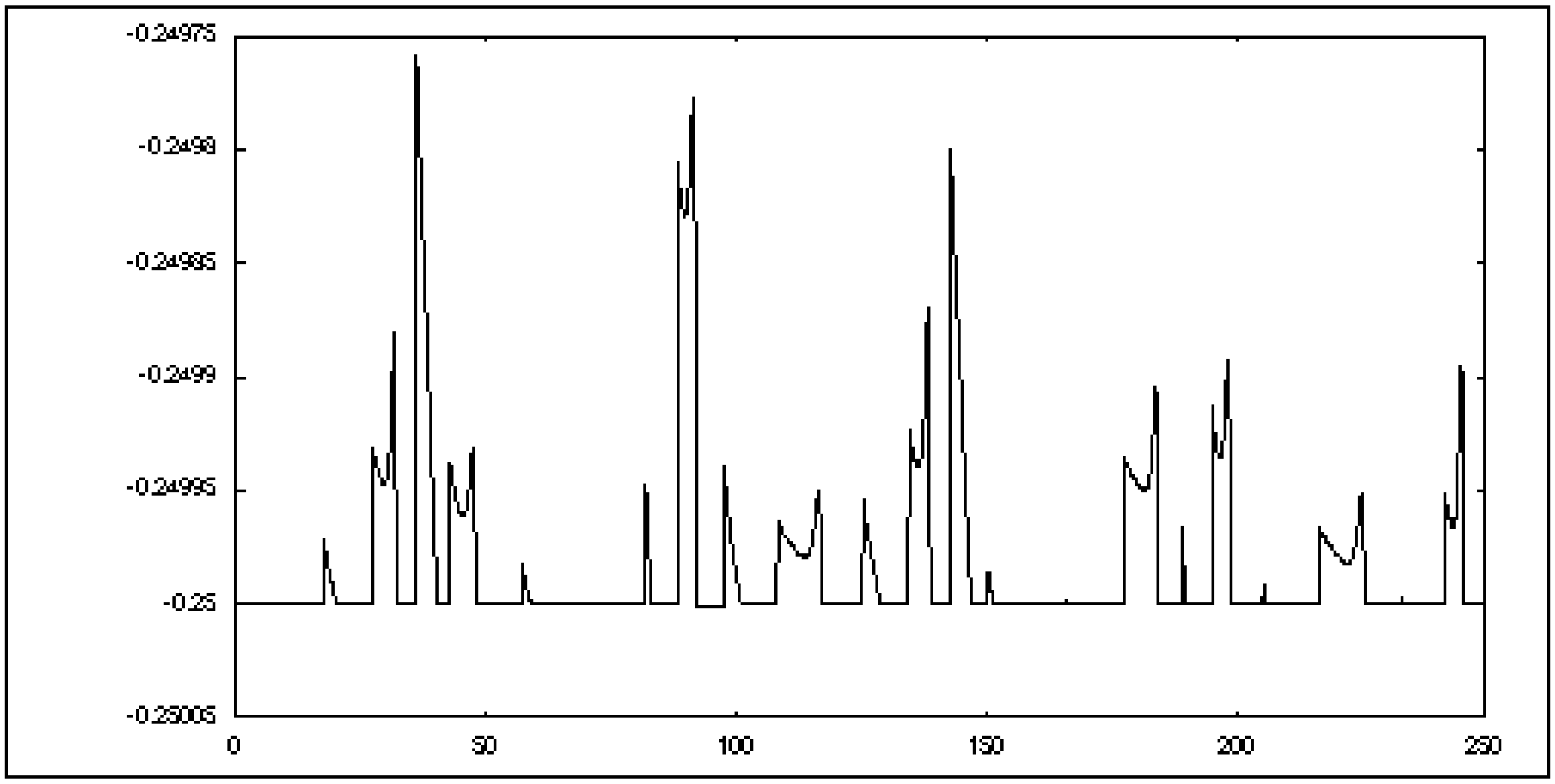

**Figure 2**. Enlarged view of the result (dotted line) shown in Figure 1.

| Time step, $\delta t$ | Mean energy | Standard Error | Fluctuation from true value |
|---|---|---|---|
| 0.1 | -0.248994002 | 0.005301927 | 0.005396458 |
| 0.05 | -0.249264118 | 0.00249823 | 0.002604327 |
| 0.01 | -0.249809787 | 0.000419863 | 0.000460936 |
| 0.005 | -0.249899569 | 0.000207661 | 0.00023067 |
| 0.001 | -0.249981339 | 3.95334E-05 | 4.37162E-05 |

**Table 1**. Nearly energy-conserving numerical integration routine for different time steps.

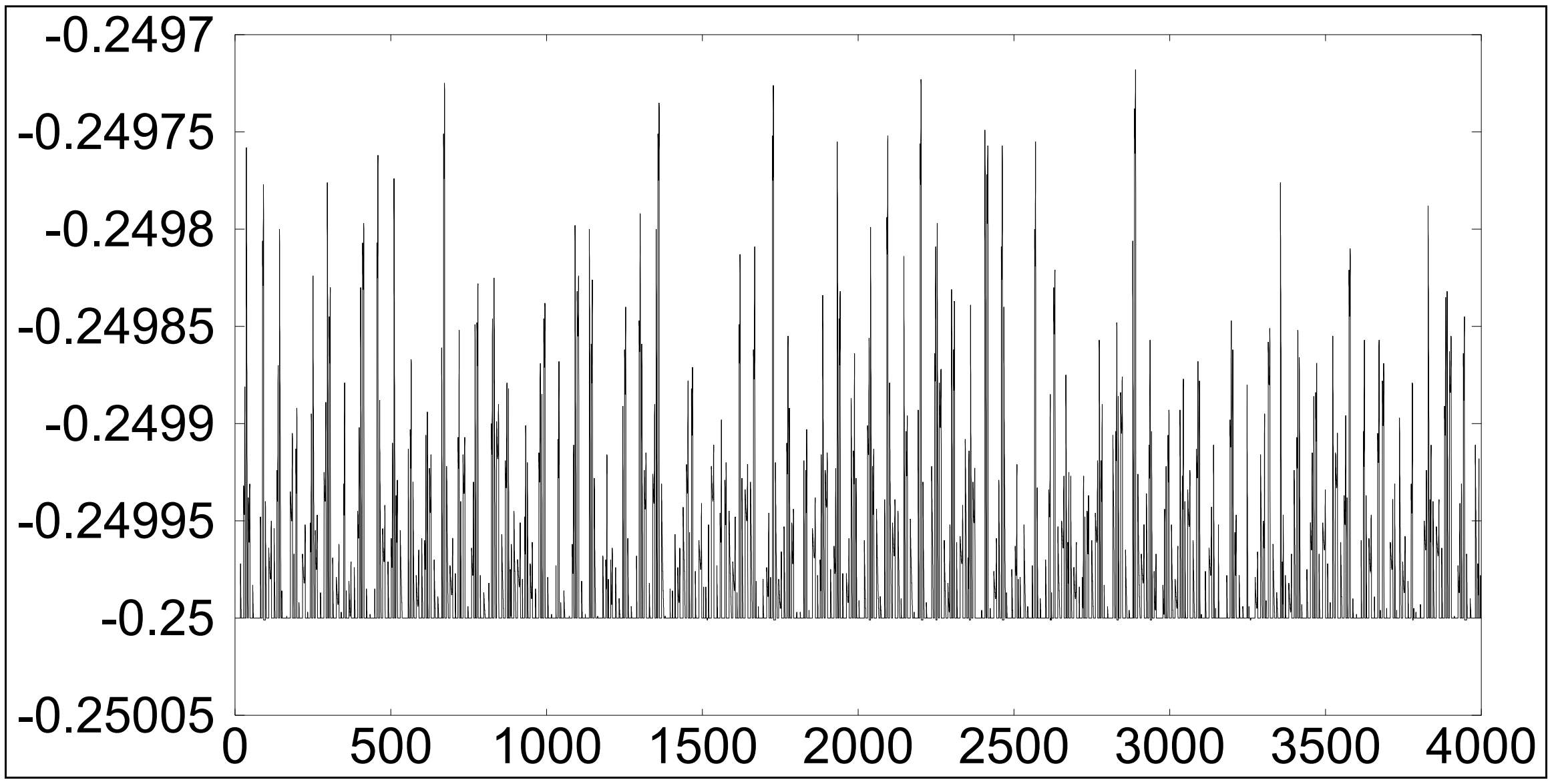

**Figure 3**. Longer time fluctuations in the total energy of a 2-shell system with $\delta t = 0.001$.

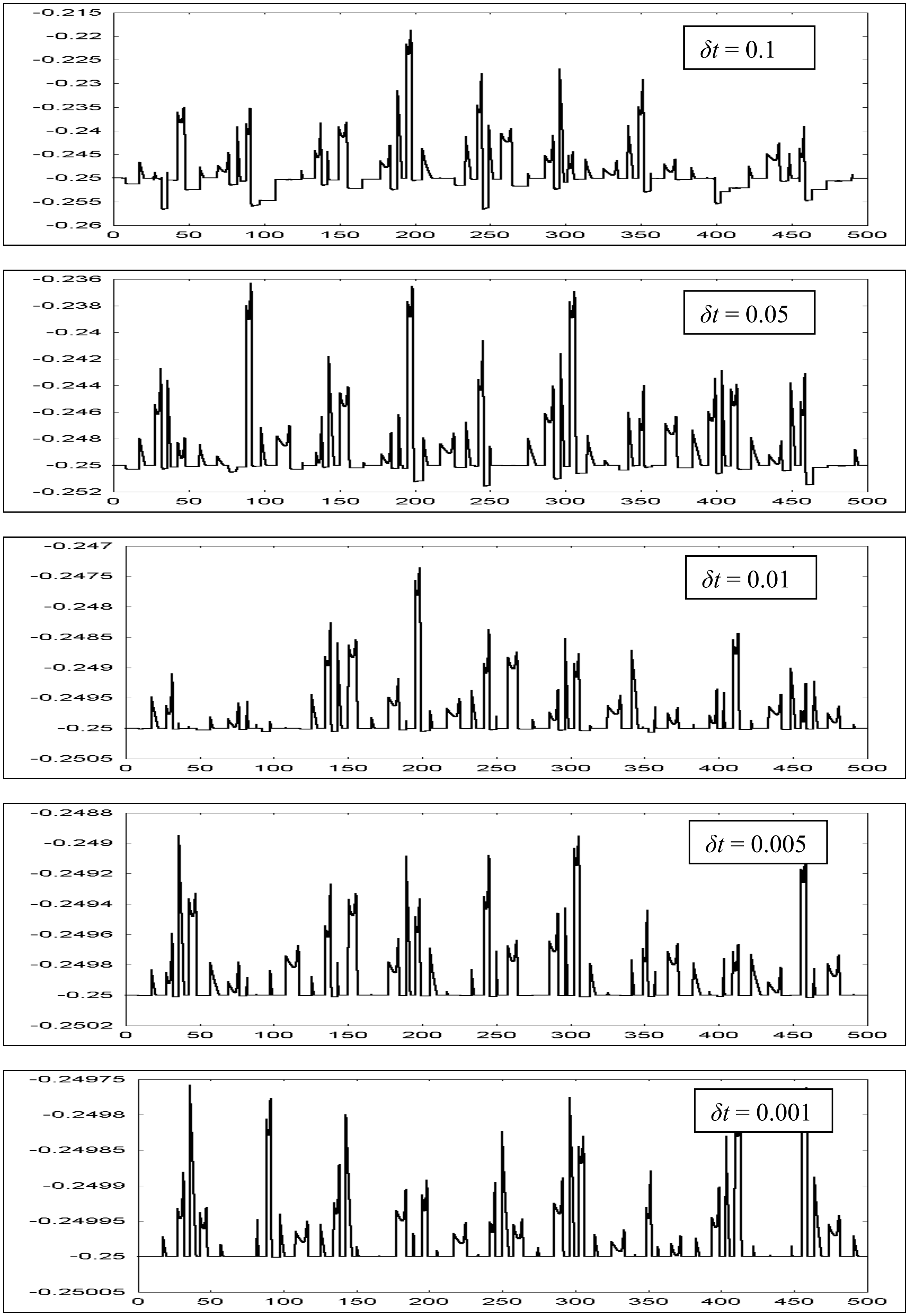

**Figure 4**. Fluctuations in the total energy of a 2-shell system for different time-step size, $\delta t$.

# Chapter 3
# Nonlinear dynamics of a rotational two-shells system

In this Chapter, we investigate the dynamical behavior of the rotational concentric 2-shells model and examine the phase space structures arising in its dynamics. In particular, we identify regions exhibiting fixed point, periodic, quasiperiodic, and chaotic dynamics by using Poincare sections, phase-space projections, power spectrum, and time series plots for the separation of two trajectories corresponding to two closely spaced initial conditions. We present the equations of motion of the system in Section 3.1 then present a method for characterizing periodic solutions in Section 3.2. In the succeeding sections, we discussed the other types of solution (i.e. quasiperiodic and chaotic solutions). Finally, we conclude the chapter with additional remarks regarding the implications of the behaviors observed in the few shell systems to the dynamics of the *N*-shell system for large *N*.

## 3.1 Equations of motion

The *N*-shell system is a 2*N*-dimensional dynamical system described by the following equations of motion

$$\frac{dr_k}{dt} = v_k \tag{3.1}$$

$$\frac{dv_k}{dt} = \frac{L^2}{m^2 r_k^{\ 3}} - \frac{Gm}{r_k^{\ 2}}\left(\frac{1}{2} + \sum_{j=1}^{N} H\left(r_k - r_j\right)\right) \tag{3.2}$$

where $(r_k, v_k)$ describes the state of shell *k*, and *k* ranges from 1 to *N*. In this chapter we deal with the $N = 2$ case. Barkov et al [40] studied this system previously. In that study their main objective was to propose a mechanism that causes the expulsion of one shell. They also obtained time series plots suggesting the occurrence of chaos for initial conditions where no shell was expelled but as the investigation of the chaotic dynamics was not the main aim of the paper, they did not perform further tests to ascertain whether the behavior was really chaotic. In this thesis, however, we only consider systems which are boundedly stable (i.e. no escape occurs). For $N = 2$, the no-escape condition is equivalent to the requirement that the system's energy is within the range $(-G^2m^5/L^2, -(1/8)G^2m^5/L^2)$.

In the following sections: we present Poincare surface of sections (3.2), chaotic and strange attractors (3.3), fixed point and periodic solutions (3.4), and quasiperiodic solutions (3.5) corresponding to the rotating 2-spherical shells system. For further discussion of this tools for analyzing dynamical systems see Nayfeh [22]. Again, the time series used for these plots were obtained using the new integrator introduced in Chapter 2 for N = 2. All programs used to generate the plots in this chapter appear in Appendix C.

## 3.2 Poincare surface of section

To visualize the system's dynamics, we construct a three-dimensional Poincare section of the four-dimensional phase-space. We use the surface corresponding to crossings between the shells for this purpose. Our Poincare section is thus given by the closed constant energy surface

$$R\left(v_1, v_2\right) = \frac{Gm^2}{E_0 - K}\left[-1 \pm \sqrt{1 + \frac{L^2\left(E_0 - K\right)}{G^2 m^5}}\right] \tag{3.3}$$

where $K = ½m(v_1^{\ 2} + v_2^{\ 2})$. Its projection in the $v_1v_2$-plane is bounded by the circle with radius equal to the maximum possible velocity of the 2-shell system. Figure 5 shows the shape of a Poincare surface of section. In this Poincare surface of section plot, the vertical axis refers to the radial position at which the shells cross while the horizontal axes refers to their crossing velocities.

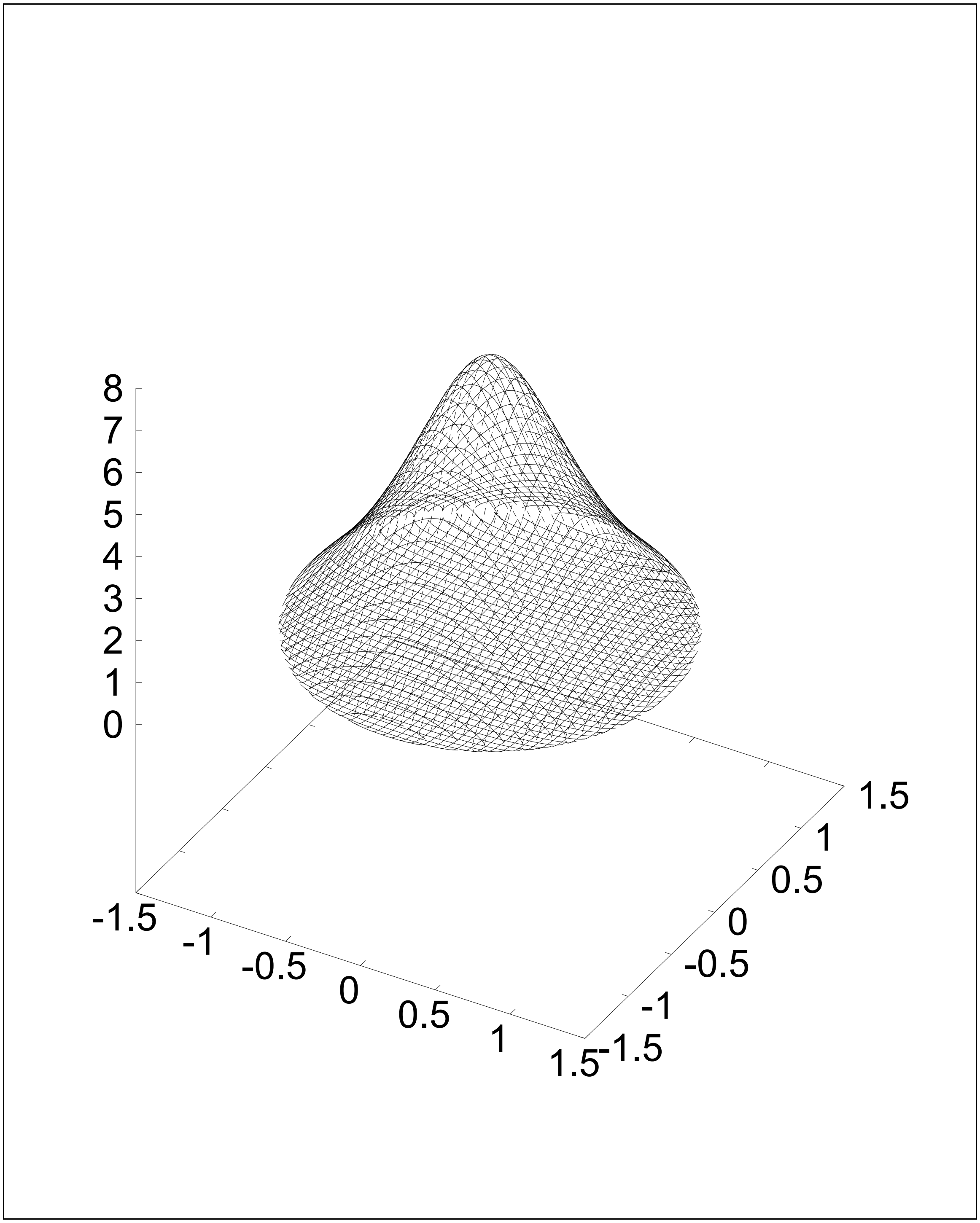

**Figure 5**. Two-sided Poincare surface of section of a 2-shell system. This is obtained by projecting the four-dimensional dynamical system to the $r_1v_1v_2$-space during crossings (i.e. $r_1 = r_2$).

## 3.3 Chaotic and strange attractors

In Figure 6, the chaotic region -- represented by irregularly scattered points -- occupies large portion of the entire Poincare surface. Its power spectrum (Figure 7) clearly shows the broadband characteristic of chaotic motion. A plot of separation distance as a function of time between two initially very close chaotic trajectories (Figure 8) strongly indicates the sensitivity to initial conditions.

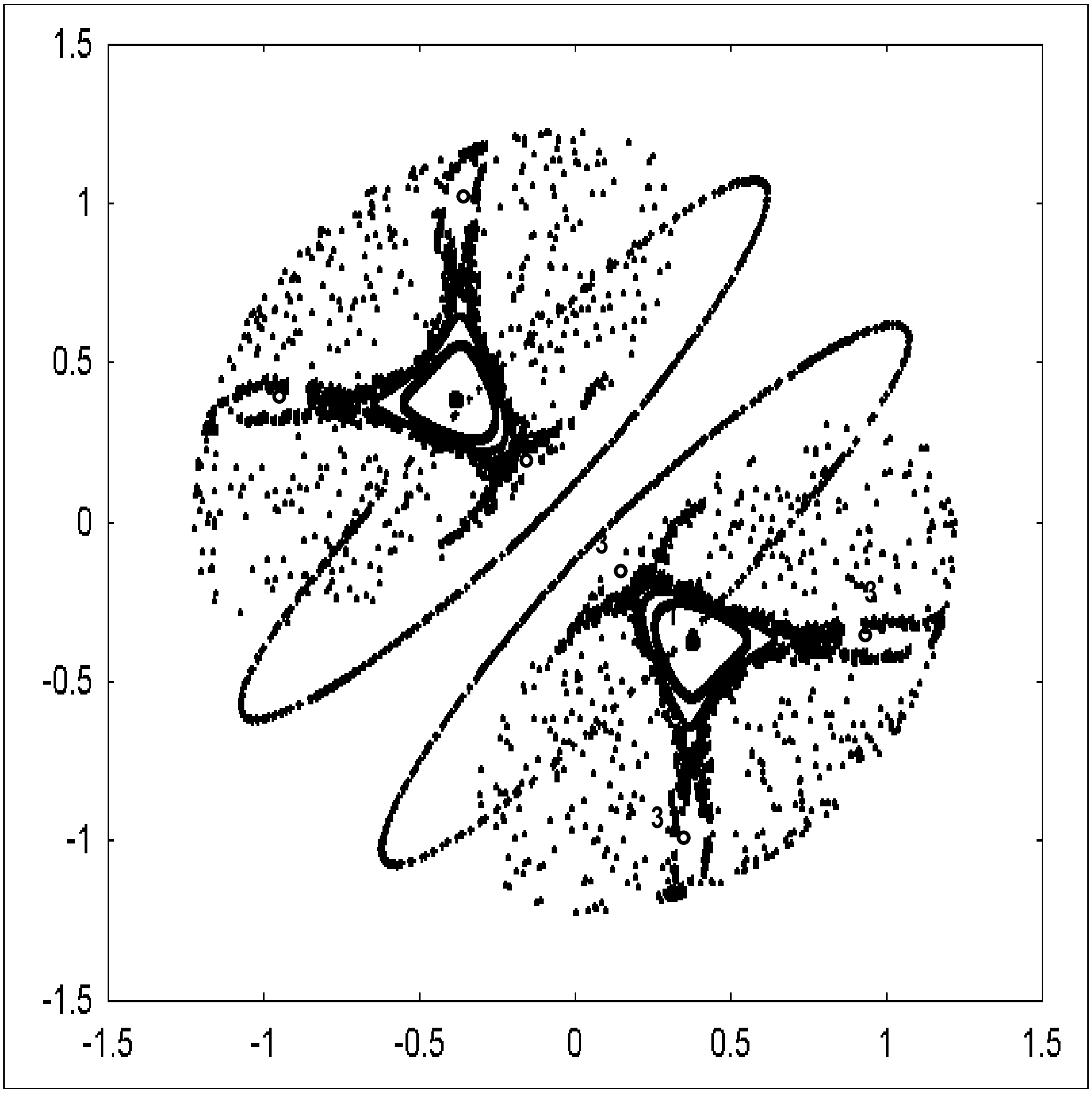

**Figure 6**. Projection at $v_1v_2$-plane of two-sided Poincare sections showing aperiodic solutions.

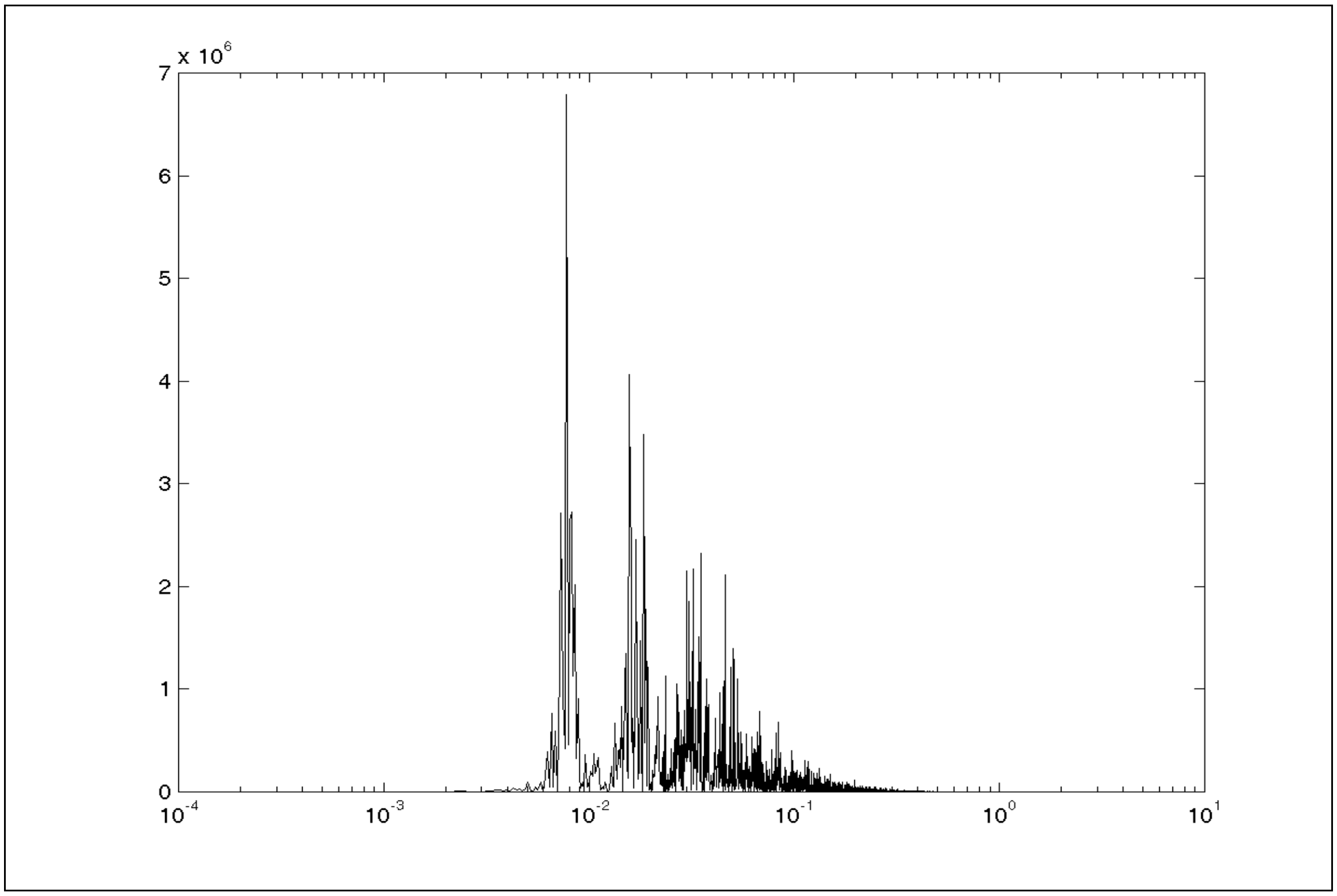

**Figure 7**. Power spectrum of $v_1$ corresponding to the chaotic region in Figure 5 consisting of widely scattered points.

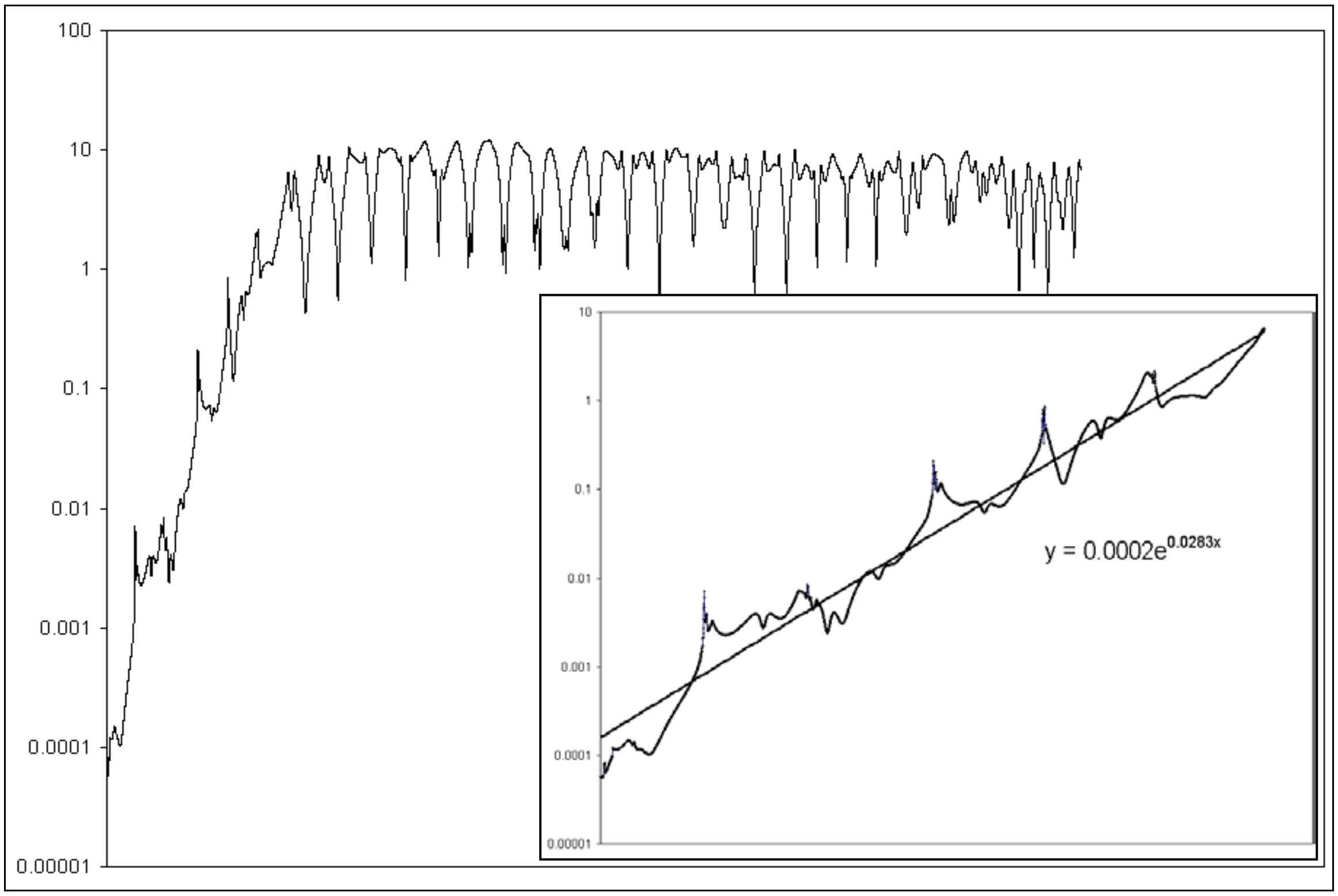

**Figure 8**. Separation distance as a function of time between two chaotic trajectories of a 2-shell system showing the sensitivity to initial condition.

## 3.4 Fixed point and periodic solutions

### 3.4.1 Fixed point and small oscillations

At the lowest possible energy, $E_{0,\min}$, the system is in a stable static equilibrium or a fixed point. At slightly higher energy, periodic solutions exist. We found it difficult to apply existing fixed point linearization methods to the original equation of motion (equations 3.1 and 3.2) because of the discrete jump in the force term. We will use instead time integrals to calculate the period at small oscillations. Assuming the shells cross at only one point in space all through out the system's evolution. Then the time integral is written as

$$\int_R^{r\max} \frac{dr}{\sqrt{\frac{2}{m}\left[E_2+\frac{3Gm^2}{2r}-\frac{L^2}{2mr^2}\right]}} = \int_{r\min}^{R} \frac{dr}{\sqrt{\frac{2}{m}\left[E_1+\frac{Gm^2}{2r}-\frac{L^2}{2mr^2}\right]}} \quad (3.4)$$

which is approximately equal to

$$\int_0^{r\max-R} \frac{dx}{\sqrt{\frac{2}{m}\left[E_2+\frac{3Gm^2}{2R}\left(1-\frac{x}{R}\right)-\frac{L^2}{2mR^2}\left(1-\frac{2x}{R}\right)\right]}}$$
$$= \int_{r\min-R}^{0} \frac{dx}{\sqrt{\frac{2}{m}\left[E_1+\frac{Gm^2}{2R}\left(1-\frac{x}{R}\right)-\frac{L^2}{2mR^2}\left(1-\frac{2x}{R}\right)\right]}} \quad (3.5)$$

The crossing radius is obtained from the above equation. With the approximation $x \ll R$, the period is approximately equal to

$$T=\frac{2\left(r_{\max}-R\right)}{Rv_R}\left[2R-\frac{r_{\max}-R}{mv_R^{\,2}}\left(\frac{L^2}{mR^2}-\frac{3Gm^2}{2R}\right)\right] \quad (3.6)$$

where

$$r_{\max}=-\frac{3Gm^2}{4E_2}\left\{1+\sqrt{1-\frac{8E_2}{9E_{0,\min}}}\right\} \quad (3.7)$$

$$E_2=\frac{1}{2}\left(E_0-\frac{Gm^2}{R}\right)$$

$$E_1=\frac{1}{2}\left(E_0+\frac{Gm^2}{R}\right).$$

### 3.4.2 Periodic solutions at higher energies

At higher energies, there are infinitely many possible periodic orbit of the system. A *k-point periodic orbit* is represented by $k$ points in a one-sided Poincare surface. For example, a Poincare section consisting of five points in each side represents a 5-point periodic solution. As suggested in Figure 6, only two coexisting periodic orbits are stable. They are the single-point and the 3-point periodic orbits. The 3-point periodic orbit is indicated by three points marked with "3". Figures 9 and 10 show the projection of these orbits in the $r_1v_1$-plane.

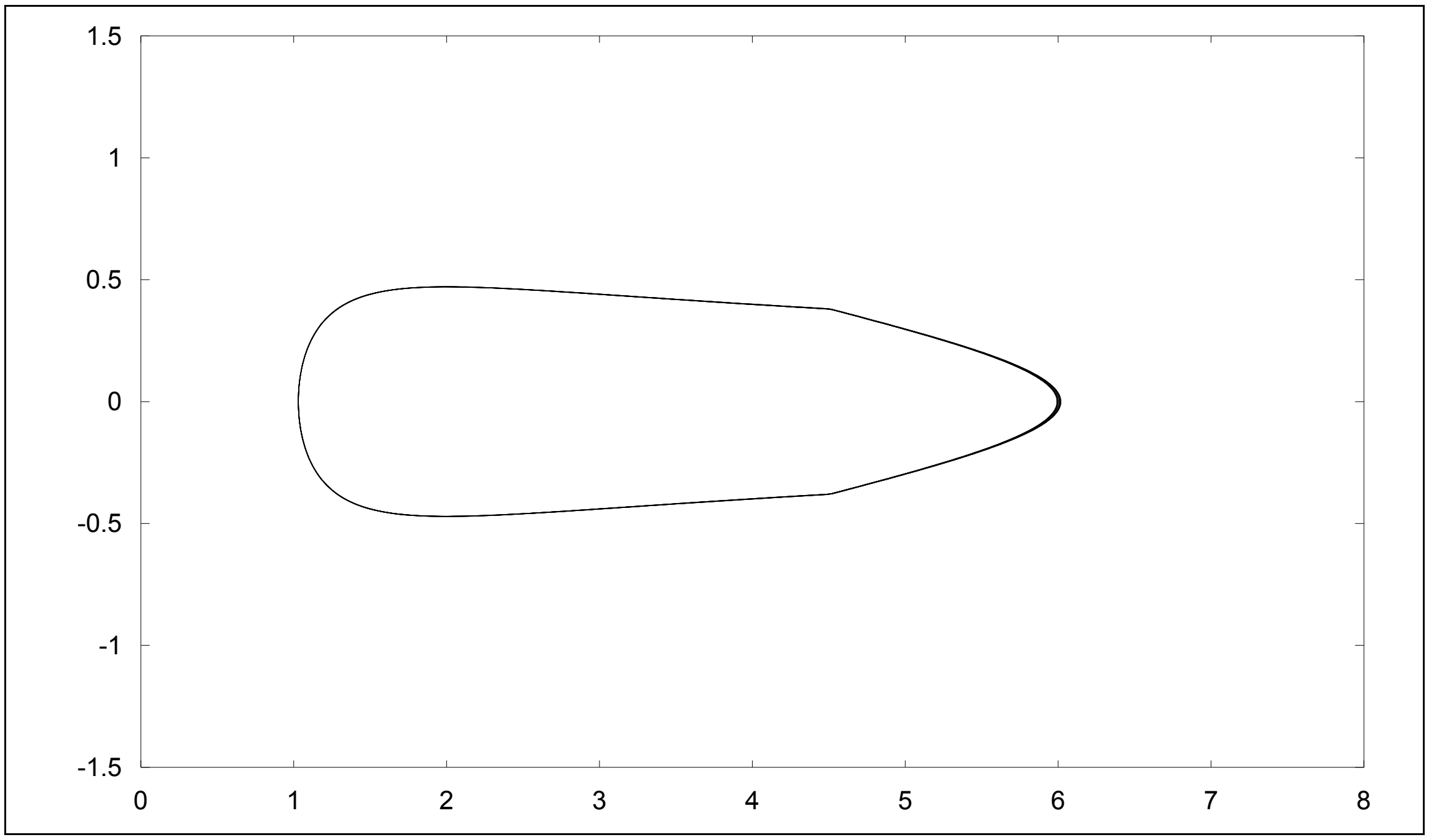

**Figure 9**. Periodic solution of a 2-shell system projected at $r_1v_1$-plane.

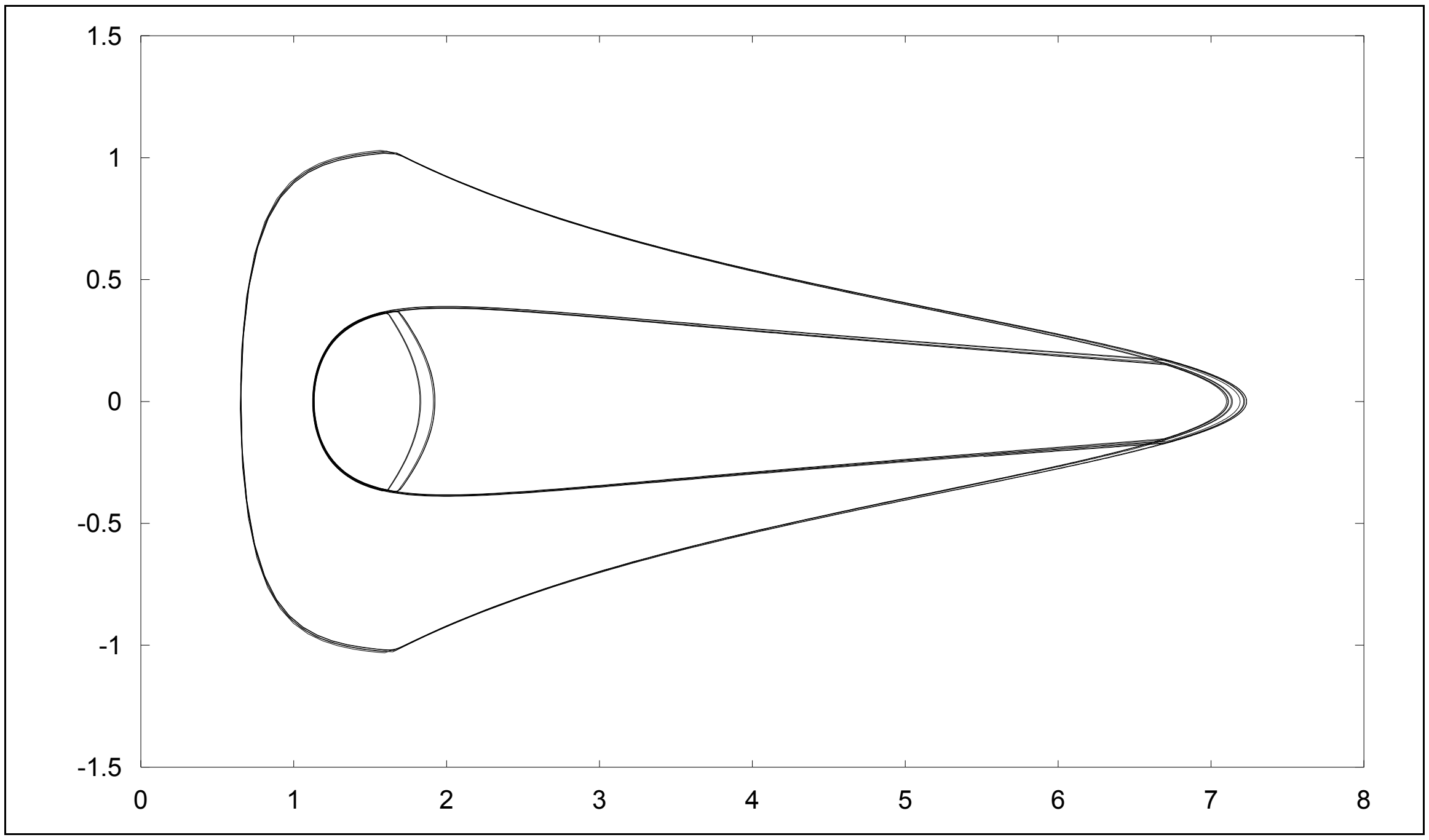

**Figure 10**. The three-point periodic solution projected at $r_1v_1$-plane.

Miller and Youngkins [13] mentioned two types of periodic orbits for the non-rotating 2-shell system. One of these orbits is analogous to the single-point periodic orbit mentioned above. This so-called *separated periodic orbit* arises from the inner shell rebounding from the origin at the same time the outer shell reaches its turning point. The shells then cross one another and the orbit is repeated with the shells crossing at exactly the same radius each time. The other periodic orbit is degenerate and occurs when the trajectories of the shells coincide. The shells act as a single particle traversing the system and rebounding at turning points. This is the *collapsed periodic orbit* and it is represented by the line $v_1 = v_2$ (not shown) in the diagram similar to Figure 6. They failed to observe the 3-point periodic orbit in the non-rotating 2-shell system, which could mean that this periodic orbit arises from a bifurcation induced by rotation.

## 3.5 Quasiperiodic solutions

A quasiperiodic orbit is characterized by two or more incommensurate frequencies—that is they can be written as superposition of two or more periodic functions with frequencies whose ratios are irrational. In the first-order Poincare surface of section, they are characterized by a dense collection of points on a closed loop. This is shown in ring-like structures surrounding the periodic orbit points in Figure 6. We may also differentiate three kinds of quasiperiodic orbit. The first one is associated with the collapsed periodic orbit and dominates the region near the $v_1 = v_2$ line. An example is shown by the oval-shaped objects in Figure 6. The other two quasiperiodic orbits enclose their respective periodic orbit points in the Poincare surface. This implies the existence of three special quasiperiodic orbits which we call *boundary orbits*. We called these boundary orbits because: one separates the collapsed quasiperiodic orbits from the chaotic region, another separates quasiperiodic orbits associated with the single-point periodic solution from those associated with the 3-point periodic orbit, and yet another one separates the non-collapsed quasiperiodic orbits from the chaotic region.

## 3.6 Additional remarks

Our numerical experiments with a 2-shell system provided us insights into the rich dynamics of the highly nonlinear system. It was already shown in past studies of parallel sheet system how stochastic behavior develops as a result of inherently large number of degrees of freedom and multi-particle collisions or intersections, which could cause discrete jumps in particle accelerations [8,42,11,12]. It has also been suggested that chaos in a 3-sheet system comes from close encounters of all three particles [24,43]. Despite these observations, parallel sheets systems do not seem to relax in a finite time (or if they relax at all, the relaxation time is much longer than current simulation times). Shells system, on the other hand, relax much more rapidly [23]. It may be possible to attribute this rapid relaxation to the highly nonlinear character of the system – two sheets systems are never chaotic but we have shown above how chaos in the rotating 2-shells system could arise.

# Chapter 4
# Time-evolution of single-particle phase-space distribution using perturbation theory

In this chapter we analytically solve for the time-evolution of the single-particle phase-space distribution function using a perturbation method adapted from Mineau et al [44] and numerically simulate the early collisionless stage of the rotational *N*-shells system's evolution. The analytical calculations start with the Vlasov-Poisson or collisionless Boltzmann equation and proceed by implementing a perturbation method to obtain approximate solution to the equation. The analytical and numerical results are then compared by constructing the number density of the shells system at a later time.

The chapter proceeds as follows: First, we discuss how to describe the early evolution of the shells system at the continuum limit (Sections 4.1); Next, we present the perturbation theoretic method of obtaining an approximate solution (Section 4.2) and implement the method for the case where the initial single-particle phase-space distribution is variable separable (Section 4.3). We then proceed by applying the result to an expanding shells system with a given initial phase-space distribution (Section 4.4) and compare the result with that of the numerical simulations (Section 4.5).

## 4.1 Early evolution of the shells system at the Vlasov limit

### The Liouville equation

Classical systems of *N* point masses described by their coordinates ($x_i$, $v_i$) can also be described in terms of an *N*-particle phase-space distribution function

$$f^{(N)}\left(\mathbf{x}_1,\mathbf{x}_2,\ldots,\mathbf{x}_N,\mathbf{v}_1,\mathbf{v}_2,\ldots,\mathbf{v}_N\right). \tag{4.1}$$

Evolution of the system is then governed by classical Liouville equation

$$\frac{\partial f^{(N)}}{\partial t}+\frac{d\mathbf{x}}{dt}\cdot\nabla f^{(N)}+\frac{d\mathbf{v}}{dt}\cdot\nabla_{\mathbf{v}} f^{(N)}=0 \tag{4.2}$$

In one-dimensional systems, this can be expanded as

$$\frac{\partial f^{(N)}}{\partial t}+\sum_{i=1}^{N}\left\{\frac{dx_i}{dt}\frac{\partial f^{(N)}}{\partial x_i}+\frac{dv_i}{dt}\frac{\partial f^{(N)}}{\partial v_i}\right\}=0 \tag{4.3}$$

The Liouville equation, however, is in practice very difficult to solve large. The *N*-particle probability density contains much more information than we would ever need or want. This is motivation enough to resort to other reduced equations such as kinetic equations encountered in the truncated BBGKY hierarchy [45].

### The collisionless Boltzmann equation

At the Vlasov limit ($N \to \infty$, $m \to 0$, Nm → constant ), the *N*-particle phase-space distribution function $f^{(N)}(x_1, x_2, \ldots, x_N, v_1, v_2, \ldots, v_N)$ can be written as the product of single-particle phase-space distribution functions $f^{(1)}(x_1, v_1) \cdots f^{(1)}(x_N, v_N)$ [46,47]. For convenience, we omit the superscript (1) and write the single-particle distribution function simply as $f(x, v)$. Thus the system is governed by the collisionless Boltzmann equation (or Vlasov equation) and the corresponding spherically symmetric Poisson equation as follows:

$$\frac{\partial f}{\partial t}+v\frac{\partial f}{\partial r}+g\left(r\right)\frac{\partial f}{\partial v}=0 \tag{4.4}$$

$$\frac{d}{dr}\left(r^2 g\left(r\right)\right)=-G\nu\left(r\right) \tag{4.5}$$

where $\nu(r)$ is the radial (linear) mass density.

## The effective gravitational field

For the rotational concentric shells system, instead of eqn(4.5) we use the following explicit expression for the effective gravitational field to apply the perturbative theory by Mineau et al [44].

$$g(r,t) = \frac{L^2}{m^2 r^3} - \frac{Gm}{r^2}\left\{\frac{1}{2} + \int_0^r n(x,t)\,dx\right\} \tag{4.6}$$

where $L$ is the magnitude of the angular momentum of each shell.

## 4.2 The perturbation theory

We start by rewriting the collisionless Boltzmann equation [eqn.(4.4)] as

$$\frac{\partial f}{\partial t} + v\frac{\partial f}{\partial r} + \lambda g(r,t)\frac{\partial f}{\partial v} = 0 \tag{4.7}$$

We expand the distribution function and the effective field in terms of zeroth-, first-, second-, ... order perturbations.

$$f(r,v,t) = f_0(r,v,t) + \lambda f_1(r,v,t) + \lambda^2 f_2(r,v,t) + \cdots \tag{4.8}$$

$$g(r,t) = g_0(r,t) + \lambda g_1(r,t) + \lambda^2 g_2(r,t) + \cdots \tag{4.9}$$

Notice that this treatment is applicable only to system of high virial ratio ($|P|/2K \ll 1$).

After substituting these to eqn.(3.7) and grouping terms of equal order in $\lambda$, we get the following set of recursive linear partial differential equations.

$$\frac{\partial f_0}{\partial t} + v\frac{\partial f_0}{\partial r} = 0$$

$$\frac{\partial f_1}{\partial t} + v\frac{\partial f_1}{\partial r} = -g_0\frac{\partial f_0}{\partial v}$$

$$\frac{\partial f_2}{\partial t} + v\frac{\partial f_2}{\partial r} = -g_1\frac{\partial f_0}{\partial v} - g_0\frac{\partial f_1}{\partial v}$$

$$\vdots$$

$$\frac{\partial f_k}{\partial t} + v\frac{\partial f_k}{\partial r} = S_{k-1}(r,v,t) \tag{4.10}$$

where

$$n_k(r,t) = \int_{-\infty}^{+\infty} f_k(r,v,t)\,dv \tag{4.11}$$

$$g_0(r,t) = \frac{L^2}{m^2 r^3} - \frac{Gm}{r^2}\left\{\frac{1}{2} + \int_0^r n_0(x,t)\,dx\right\} \tag{4.12a}$$

$$g_k(r,t) = -\frac{Gm}{r^2}\int_0^r n_k(x,t)\,dx, \quad k > 0 \tag{4.12b}$$

$$S_k(r,v,t) = -\sum_{j=0}^{k} g_{k-j}\frac{\partial f_j}{\partial v} \tag{4.13}$$

The recursive method in approximating successive orders in the perturbative expansion [eqns(4.10)] works as follows: Given the initial phase-space distribution function $f(r, v, t = 0)$, the zeroth order function is $f_0(r, v, t) = f(r - vt, v, t = 0)$, which is just the description for free expansion of non-interacting gas. Then after calculating the quantities (4.11), (4.12) and (4.13) the successive orders are

$$f_k(r,v,t) = \int_0^t S_{k-1}(r - vt + vs, v, s)\,ds \tag{4.14}$$

While the method is straightforward, one discovers after following the calculations that several calculational difficulties are inherent in the nonequilibrium statistical mechanics. As presented in the next section, the first-order function is already complicated such that we content ourselves with at least the second-order approximation. So far no mathematical trick is known to simplify calculations that is why we do it brute-force.

## 4.3 Variable separable initial phase-space density

In this section, we implement the perturbation theory discussed above to the case when the initial phase-space distribution is variable separable. That is,

$$f\left(r,v,t=0\right)=n\left(r\right)\varphi\left(v\right). \tag{4.15}$$

We proceed to calculating the perturbation terms in *f*(*r*, *v*, *t*) up to first-order. We do not calculate anymore the second-order term for reasons that will be discussed later in this section.

### 4.3.1 Zeroth-order perturbation

As mentioned in section 4.4, the zeroth-order perturbation to the time-evolved phase-space distribution is written as

$$f_0\left(r,v,t\right)=f\left(r-vt,v,0\right) \tag{4.16}$$

which, in the variable separable case, is equal to

$$f_0\left(r,v,t\right)=n\left(r-vt\right)\varphi\left(v\right). \tag{4.17}$$

**Series expansion of *n*(*r* – *vt*)**

We find it useful to expand $n(r-vt)$ in eqn (4.17) as Taylor series in $v$. We express it as

$$n(r-vt)=\sum_{k=0}^{\infty}\frac{D_r^{\,k}\left[n(r)\right]}{k!}(-t)^k v^k \tag{4.18}$$

so that $f_0$ can then be written in the following form

$$f_0\left(r,v,t\right)=\sum_{k=0}^{\infty}\frac{D_r^{\,k}\left[n\left(r\right)\right]}{k!}\left(-t\right)^k v^k\phi\left(v\right). \tag{4.19}$$

**Zeroth-order number densities**

Thus in general, the initially separable distribution respectively gives the zeroth-order space and velocity densities

$$n_0\left(r,t\right)=\int_{-\infty}^{\infty} f_0\left(r,u,t\right)du=\int_{-\infty}^{\infty} n\left(r-ut\right)\phi\left(u\right)du \tag{4.20}$$

$$\phi_0\left(v,t\right)=\int_0^{\infty} f_0\left(x,v,t\right)dx=\phi\left(v\right)\int_0^{\infty} n\left(x-vt\right)dx \tag{4.21}$$

**Zeroth-order fields**

From these results, we obtain the zeroth-order perturbation for gravitational field as

$$g_0\left(r,t\right)=\frac{L^2}{m^2r^3}-\frac{Gm}{2r^2}-\frac{Gm}{r^2}\int_0^r n_0\left(x,t\right)dx$$

$$=\frac{L^2}{m^2r^3}-\frac{Gm}{2r^2}-\frac{Gm}{r^2}\int_0^r\left[\int_{-\infty}^{\infty} n\left(x-ut\right)\phi\left(u\right)du\right]dx \tag{4.22}$$

and the derivative of $f_0$ with respect to $v$ as

$$\frac{\partial f_0}{\partial v}=\frac{\partial}{\partial v}\left\{n\left(r-vt\right)\phi\left(v\right)\right\}=n\left(r-vt\right)\frac{\partial\phi}{\partial v}-t\frac{\partial n}{\partial\left(r-vt\right)}\phi\left(v\right). \tag{4.23}$$

### 4.3.2 First-order perturbation

The first-order perturbation to the phase-space density is then given by eqn (4.14)

$$f_1(r,v,t)=\int_0^t S_0(r-vt+vs,v,s)\,ds \tag{4.24}$$

where $S_0$, according to eqn (4.13), is equal to

$$S_0(r,v,t)=-g_0(r,t)\frac{\partial f_0}{\partial v}\,. \tag{4.25}$$

Since $\frac{\partial f_0}{\partial v}$, after the transformations $t \rightarrow s$ and $r \rightarrow r - vt + vs$, has two terms in $s$ and $g_0$ has two plus one integral, we effectively need to calculate four plus two integrals in $s$ to solve for $f_1$. We proceed as follows

$$S_0(r-vt+vs,v,s)=$$

$$-\left[\frac{L^2}{m^2(r-vt+vs)^3}-\frac{Gm}{2(r-vt+vs)^2}\right]\begin{Bmatrix} n(r-vt)\dfrac{\partial\phi}{\partial v} \\ -s\dfrac{\partial n}{\partial(r-vt)}\phi(v)\end{Bmatrix}$$

$$+\frac{Gm}{(r-vt+vs)^2}\begin{Bmatrix} n(r-vt)\dfrac{\partial\phi}{\partial v} \\ -s\dfrac{\partial n}{\partial(r-vt)}\phi(v)\end{Bmatrix}\int_0^{r-vt+vs}\left[\int_{-\infty}^{\infty} n(x-us)\phi(u)\,du\right]dx \tag{4.26}$$

$$f_1(r,v,t)=$$

$$-n(r-vt)\frac{\partial\phi}{\partial v}\left[\frac{L^2}{2m^2r(r-vt)}\left(\frac{1}{r}+\frac{1}{r-vt}\right)-\frac{Gm}{2r(r-vt)}\right]t$$

$$+\frac{\partial n}{\partial(r-vt)}\phi(v)\left[\frac{L^2}{2m^2r^2(r-vt)}t^2-\frac{Gm}{2rv}t+\frac{Gm}{2v^2}\ln\left|\frac{r}{r-vt}\right|\right]$$

$$+n(r-vt)\frac{\partial\phi}{\partial v}Gm\int_{-\infty}^{\infty}\frac{\phi(u)\,du}{u}\sum_{k=0}^{\infty}\frac{(-1)^k}{(k+1)!}\left[\begin{matrix}\displaystyle\sum_{j=1}^{k}\left(-\frac{v}{u}\right)^{k-j}\frac{(k-1)!}{(j-1)!} \\ \times\left[\begin{matrix} n^{(j-1)}(ut)r^{j-1} \\ -n^{(j-1)}(0)(r-vt)^{j-1}\end{matrix}\right]\end{matrix}\right]$$

$$-\frac{\partial n}{\partial(r-vt)}\phi(v)Gm\int_{-\infty}^{\infty}\frac{\phi(u)\,du}{u^2}\sum_{k=0}^{\infty}\frac{(-1)^k}{(k+1)!}\left[\begin{matrix}\displaystyle\sum_{j=1}^{k}\left(-\frac{v}{u}\right)^{k-j}\frac{(k-1)!}{(j-1)!} \\ \times\begin{Bmatrix}\left[\begin{matrix} utn^{(j-1)}(ut) \\ -(k-j+1)n^{(j-2)}(ut)\end{matrix}\right]r^{j-1} \\ +(k-j+1)n^{(j-2)}(0)(r-vt)^{j-1}\end{Bmatrix}\end{matrix}\right] \tag{4.27}$$

We now expand the summation terms appearing in eqn.(3.27) as follows

$$\sum_{k=0}^{\infty}\frac{(-1)^k}{(k+1)!}\left[\begin{array}{l}\displaystyle\sum_{j=1}^{k}\left(-\frac{v}{u}\right)^{k-j}\frac{(k-1)!}{(j-1)!}\\ \times\left[\begin{array}{l} n^{(j-1)}(ut)r^{j-1}\\ -n^{(j-1)}(0)(r-vt)^{j-1}\end{array}\right]\end{array}\right]=\sum_{k=0}^{\infty}\frac{(-1)^k}{(k+1)!}A_k(r,v,t,u) \qquad (4.28)$$

$$\sum_{k=0}^{\infty}\frac{(-1)^k}{(k+1)!}\left[\begin{array}{l}\displaystyle\sum_{j=1}^{k}\left(-\frac{v}{u}\right)^{k-j}\frac{(k-1)!}{(j-1)!}\\ \times\left\{\begin{array}{l}\left[\begin{array}{l} utn^{(j-1)}(ut)\\ -(k-j+1)n^{(j-2)}(ut)\end{array}\right]r^{j-1}\\ +(k-j+1)n^{(j-2)}(0)(r-vt)^{j-1}\end{array}\right\}\end{array}\right]=\sum_{k=0}^{\infty}\frac{(-1)^k}{(k+1)!}B_k(r,v,t,u) \qquad (4.29)$$

where the $A_k$ and $B_k$ are listed in Appendix B. We may then simplify eqns. (4.28) and (4.29) by substituting the expressions for $A_k$s and $B_k$s. We must point out that this method may not give a closed form for first-order perturbation theory. This possibility depends on the property of the initial phase-space distribution function. In this case, we would truncate the series up to some order -- a process that would effectively reduce the time when the prediction is valid.

There is special case that would simplify the expression for $f_1$. When $n(r) = C[\delta(r - z) + \delta(r + z)]$, its derivative to any order is zero at $r = 0$. In the next section we will discuss this particular case of expanding spherical shells.

## 4.4 Expanding spherical shells

### 4.4.1 Initial condition

We study the phase-space evolution of an expanding spherical shell described initially by the number densities (see also Figure 11)

$$n(r)=\lim_{b\to\infty}\sqrt{\frac{b}{\pi}}\left\{\exp\left[-b(r-z)^2\right]+\exp\left[-b(r+z)^2\right]\right\} \qquad (4.30)$$

$$\varphi(v)=\lim_{b\to\infty}\frac{1}{4s}\left\{\tanh\left[b(v+s)\right]-\tanh\left[b(v-s)\right]\right\} \qquad (4.31)$$

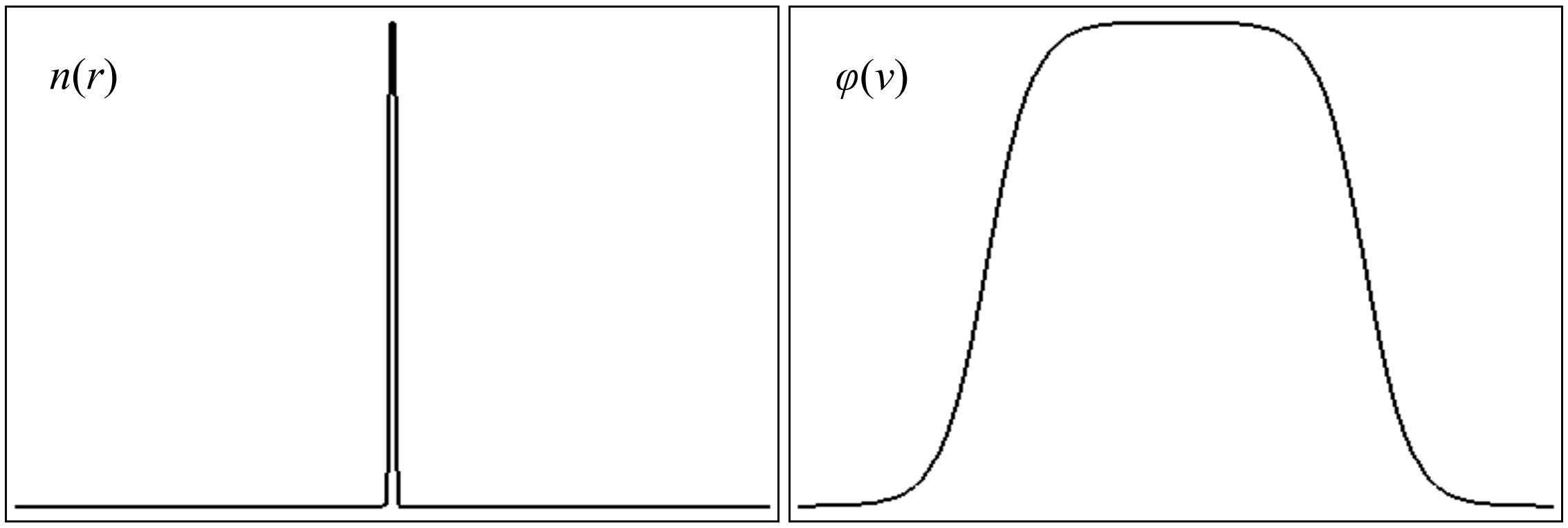

**Figure 11**. Initial phase-space distribution $f(r, v, 0) = n(r)\varphi(v)$.

### 4.4.2 Analytical treatment

For this particular example, we calculate the functions $A_k$ and $B_k$ (see Appendix B) and write $f_1$ as

$$f_1(r,v,t) =$$

$$-n(r-vt)\frac{\partial\phi}{\partial v}\left[\frac{L^2}{2m^2 r(r-vt)}\left(\frac{1}{r}+\frac{1}{r-vt}\right)-\frac{Gm}{2r(r-vt)}\right]t$$

$$+n^{(1)}(r-vt)\phi(v)\left[\frac{L^2}{2m^2r^2(r-vt)}t^2-\frac{Gm}{2rv}t+\frac{Gm}{2v^2}\ln\left|\frac{r}{r-vt}\right|\right]$$

$$+n(r-vt)\frac{\partial\phi}{\partial v}\frac{Gm}{2s}\int_{-s}^{s}\frac{du}{u}\left[A_0-\frac{A_1}{2!}+\sum_{k=2}^{\infty}\frac{(-1)^k}{(k+1)!}A_k\right]$$

$$-n^{(1)}(r-vt)\phi(v)\frac{Gm}{2s}\int_{-s}^{s}\frac{du}{u^2}\left[B_0-\frac{B_1}{2!}+\frac{B_2}{3!}+\sum_{k=3}^{\infty}\frac{(-1)^k}{(k+1)!}B_k\right]. \qquad (4.32)$$

By symmetry, only $A_0$, $B_0$, and $B_2$ contribute to the first series[$].

$$I_0=\int_{-s}^{s}\frac{du}{u}A_0=2\int_0^s\frac{H(ut-z)du}{u(r-vt)+vz}=\frac{2}{(r-vt)}\ln\left|\frac{t\left[(r-vt)+vT\right]}{rT}\right|H(t-T)$$

$$J_0=\int_{-s}^{s}\frac{du}{u^2}[B_0]=2zt\int_0^{st}\frac{H(x-z)dx}{x^2(r-vt)+xtvz}=4\left[\frac{t}{2(r-vt)-vt}-\frac{T}{2(r-vt)-vT}\right]H(t-T)$$

$$J_2=\int_{-s}^{s}\frac{du}{u^2}\left[\frac{B_2}{3!}\right]=-\frac{1}{3}\left[\frac{vT^2+rt}{z^2}\right]H(t-T)$$

Using only the first two terms in the series, we obtain an approximation to the first-order perturbation of the phase-space density given by

$$f(r,v,t)\approx f_0(r,v,t)+f_1(r,v,t)\approx$$

$$n(r-vt)\varphi(v)$$

$$-n(r-vt)\frac{\partial\phi}{\partial v}\left[\frac{L^2}{2m^2 r(r-vt)}\left(\frac{1}{r}+\frac{1}{r-vt}\right)-\frac{Gm}{2r(r-vt)}\right]t$$

$$+n^{(1)}(r-vt)\phi(v)\left[\frac{L^2}{2m^2r^2(r-vt)}t^2-\frac{Gm}{2rv}t+\frac{Gm}{2v^2}\ln\left|\frac{r}{r-vt}\right|\right]$$

$$+n(r-vt)\frac{\partial\phi}{\partial v}\frac{Gm}{s(r-vt)}\ln\left|\frac{t\left[(r-vt)+vT\right]}{rT}\right|H(t-T)$$

$$-n^{(1)}(r-vt)\phi(v)\frac{Gm}{s}\left[\frac{2t}{2(r-vt)-vt}-\frac{2T}{2(r-vt)-vT}-\frac{vT^2+rt}{6z^2}\right]H(t-T)$$

For the early part ($t < T$) of the evolution, the number density is given by (See also equation 4.34)

$$n_1(r)=\int_{-\infty}^{\infty}f(r,v,t)dv \qquad (4.33)$$

$$=\frac{1}{t}\left[\varphi\left(\frac{r+z}{t}\right)+\varphi\left(\frac{r-z}{t}\right)\right]$$

[$] $T = z/s$.

$$-\left\{\varphi^{(1)}\left(\frac{r+z}{t}\right)\left[\frac{L^2}{2m^2rz}\left(\frac{r-z}{rz}\right)+\frac{Gm}{2rz}\right]+\varphi^{(1)}\left(\frac{r-z}{t}\right)\left[\frac{L^2}{2m^2rz}\left(\frac{r+z}{rz}\right)-\frac{Gm}{2rz}\right]\right\}$$

$$+\left\{\begin{array}{l}\frac{L^2}{2m^2r^2z}\left[\varphi^{(1)}\left(\frac{r-z}{t}\right)-\varphi^{(1)}\left(\frac{r+z}{t}\right)\right]+\frac{Gm}{2}\ln\left|\frac{r}{z}\right|\left[\frac{\varphi^{(1)}\left(\frac{r-z}{t}\right)}{(r-z)^2}+\frac{\varphi^{(1)}\left(\frac{r+z}{t}\right)}{(r+z)^2}\right]\\ -\frac{Gm}{2}\left[\frac{\varphi^{(1)}\left(\frac{r-z}{t}\right)}{r(r-z)}+\frac{\varphi^{(1)}\left(\frac{r+z}{t}\right)}{r(r+z)}\right]\end{array}\right\}$$

$$+\left\{\begin{array}{l}t\frac{L^2}{2m^2r^2z^2}\left[\varphi\left(\frac{r+z}{t}\right)+\varphi\left(\frac{r-z}{t}\right)\right]-Gmt\ln\left|\frac{r}{z}\right|\left\{\frac{\varphi\left(\frac{r-z}{t}\right)}{(r-z)^3}+\frac{\varphi\left(\frac{r+z}{t}\right)}{(r+z)^3}\right\}\\ +\frac{Gmt}{2r}\left[\frac{\varphi\left(\frac{r-z}{t}\right)}{(r-z)^2}+\frac{\varphi\left(\frac{r+z}{t}\right)}{(r+z)^2}\right]+\frac{Gmt}{2z}\left\{\frac{\varphi\left(\frac{r-z}{t}\right)}{(r-z)^2}-\frac{\varphi\left(\frac{r+z}{t}\right)}{(r+z)^2}\right\}\end{array}\right\}$$

## 4.5 Results of the numerical simulation

For the short time evolution of $2^n$-shells system ($n$ = 2, 3, …, 10) the results of the Verlet and the hybrid Verlet-modified Euler-Cromer schemes are identical if the evolution time is so short that there are no collisions. The absence of collisions leads to minimal fluctuations in the computed total energy and hence there is no need to switch from the Verlet to modified Euler-Cromer integration during the simulation.

In the simulations, the expanding spherical shells system as described by equations (4.30) and (4.31) is initialized in such a way that all shells have the same initial radius (= $z$) but with velocities assigned as follows: first, the shells are labeled $i$ from 1 to $N$; then the velocities varies linearly with $i$ such that the shells 1 and $N$ are moving with velocities $-s$ and $+s$, respectively. To evolve the system, we then follow the prescription that the number of shells inside shell $i$ is $i-1$, which is the case after an infinitesimally small period.

Suppose the system initially has total energy $E$ and virial ratio $VR = 1/\lambda$. Then the potential and kinetic energies of the system are given by

$$P=-\frac{GM^2}{2z}=\frac{2\lambda E}{2\lambda-1}$$

and

$$K=\frac{1}{6}\left(\frac{N+1}{N-1}\right)Ms^2+\frac{L_t^2}{2Mz^2}$$

where $M = Nm$, the total mass, and $L_t = NL$ are constants independent on $N$. With this initial condition (that is, given $E$ and $\lambda$), the initial phase-space distribution is described by the quantities

$$z=-\frac{GM^2}{2P}$$

and

$$s = \sqrt{\frac{3GM}{2z}\frac{N-1}{N+1}\left(\frac{1}{\lambda}-\frac{2L_t^2}{GM^3 z}\right)}.$$

Notice that the $N$-shell system described above approaches the irrotational case at the Vlasov limit. That is, $L = L_t/N \to 0$ as $N \to \infty$. Also, we have $m \to 0$ at the Vlasov limit. This means that, instead of equation (4.33), the number density at the Vlasov limit is approximately equal to

$$n_1(r) = \left(\frac{1}{t}+\frac{tL_t^2}{2M^2r^2z^2}\right)\left[\varphi\left(\frac{r+z}{t}\right)+\varphi\left(\frac{r-z}{t}\right)\right] \qquad (4.34)$$
$$-\frac{L_t^2}{2M^2r^2z}\left[\frac{r-2z}{z}\varphi^{(1)}\left(\frac{r+z}{t}\right)+\frac{r+2z}{z}\varphi^{(1)}\left(\frac{r-z}{t}\right)\right]$$

The plot of $n_1(r)$ when $t = T = z/s$ is shown in Figure 13.

Figure 14 shows the result of the numerical simulation using $N$ = 32, 64, 128, 256, 512, and 1024. Notice the similarity with the analytical approximation except for normalization and difference in smoothness of the curve. The latter is attributed to the graininess effect due to the finite size of the system.

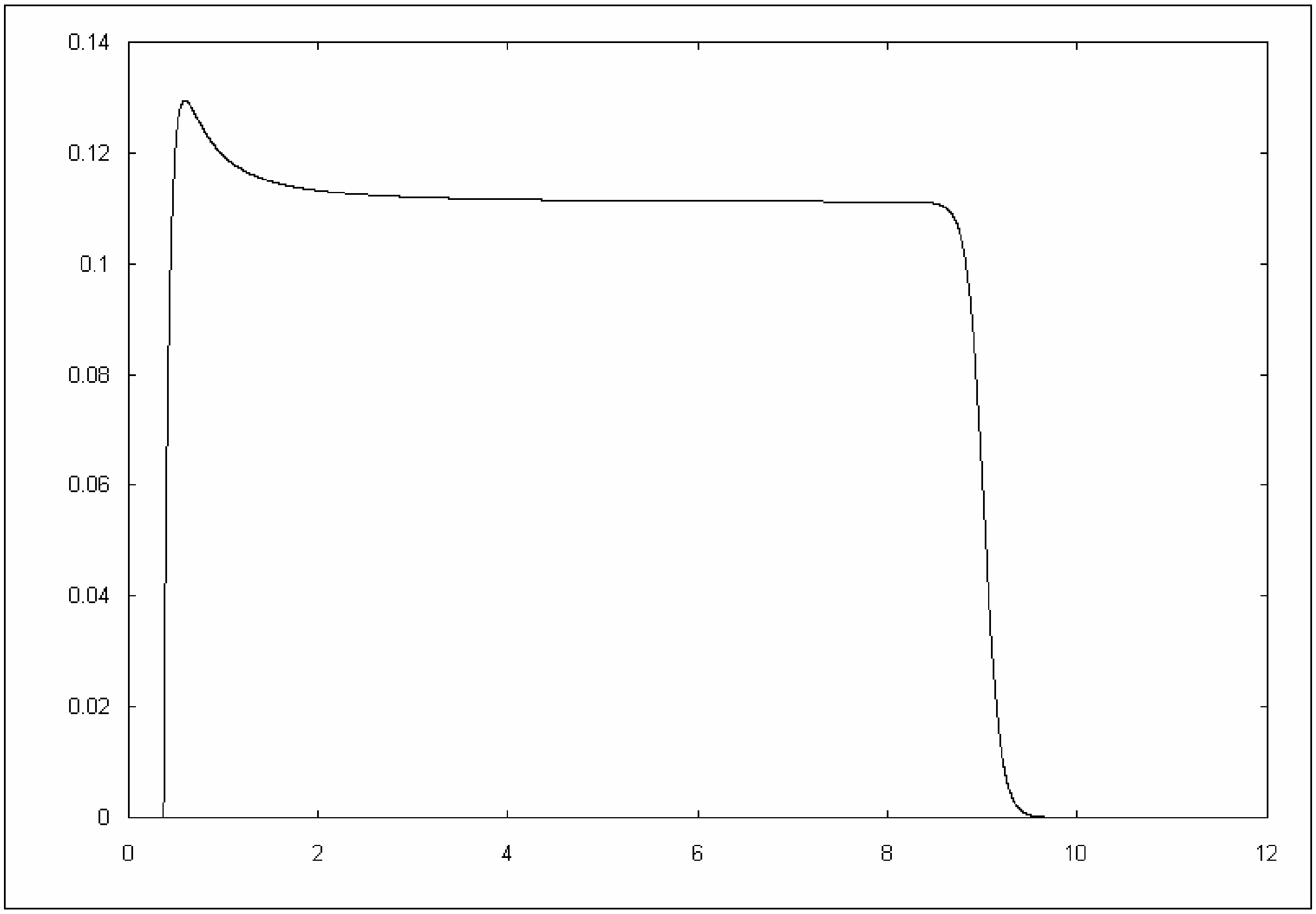

**Figure 12**. Plot of number density given by equation 4.34.

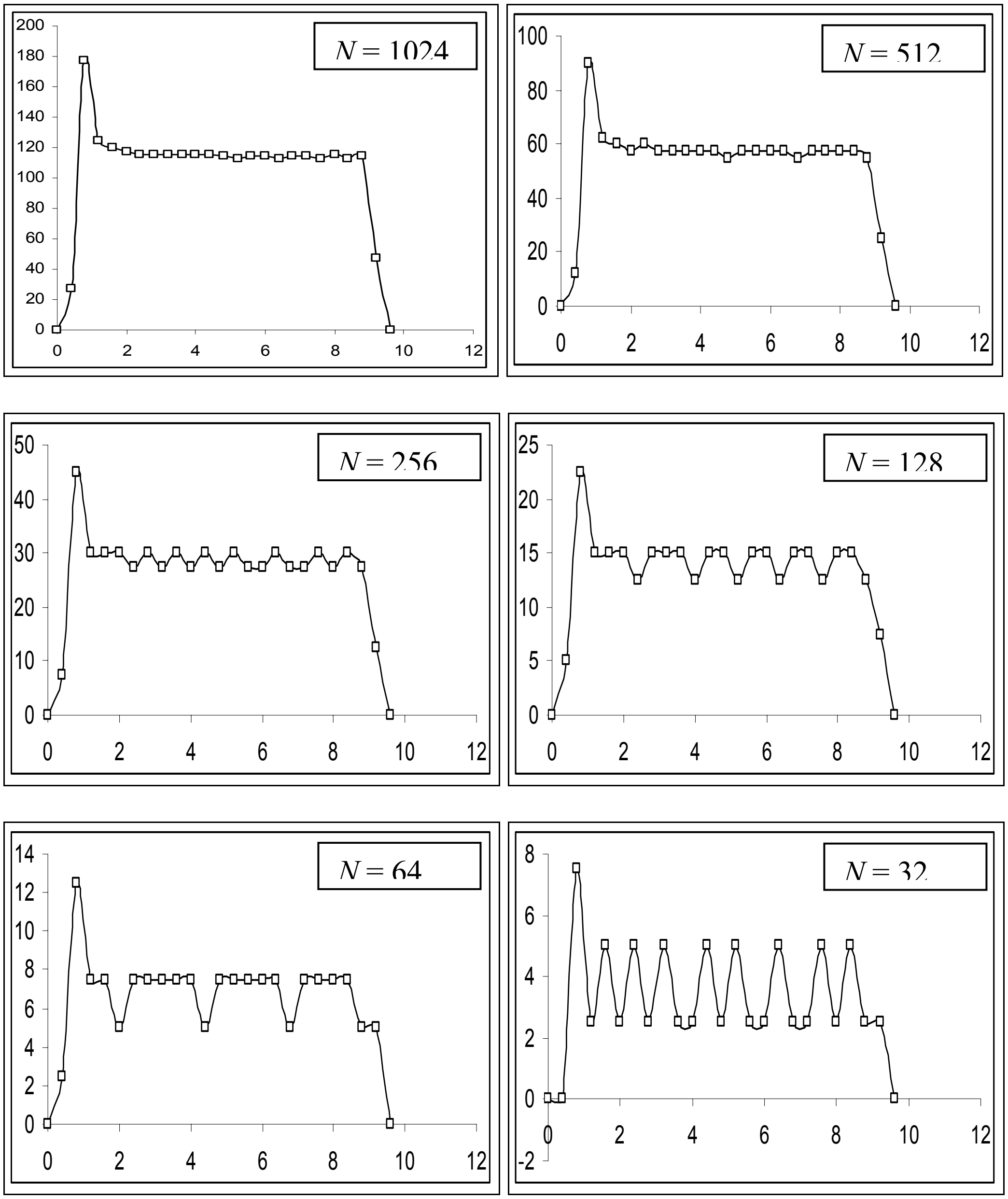

**Figure 13**. Number density at $t = T = z/s$ of $N$-shell systems initially ($t = 0$) described by equations 4.30 and 4.31 and with virial ratio 20. Notice similarity with analytical result (Figure 12).

# Chapter 5
# Conclusion

In this thesis we proposed and implemented a nearly energy conserving numerical integration scheme, we performed a detailed study of the nonlinear dynamics of the rotational 2-shell systems, and we studied the non-equilibrium behavior of the rotational spherical shells system using perturbation theory. We believe these are all new.

In order to study gravitating shells system, we proposed a numerical integration method that minimizes fluctuations in the total energy of the system. The proposed numerical integration scheme is a nearly energy conserving hybrid of the Verlet and modified Euler-Cromer integration schemes. The superior performance of the new scheme was exhibited through total energy versus time plots obtained from the new scheme and the following integrators: Euler, Euler-Cromer, modified Euler-Cromer, Verlet, fourth-order Runge-Kutta. The energy conservation requirement is critical for long-time simulations.

The new numerical integration scheme was applied to rotational two-spherical shells system. Using this method, we have successfully regulated the total energy of the system and obtained a nearly conserved energy—making possible a diagnostic study of phase-space trajectories of the system.

In the rotational 2-shell system, we observed large regions of chaotic motion. It is well known that stochasticity in higher dimensions (i.e. larger population $N$) springs from the chaotic nature of each particle's trajectory. Our assertion of chaos in this system is not new. However, the assertion in a previous study [40] failed to provide conclusive evidence for such behavior. Qualitative description of time-series plots is not enough to conclude that the trajectory is chaotic or not. In our study, we used the phase space projections, Poincare sections, power spectrum analysis, and time series plots of phase space separation distance to analyze the dynamics of the system and confirmed that is exhibits chaos.

By constructing the Poincare section of different trajectories of the rotational 2-shells system, we were able to reveal another periodic solution aside from the analogue of the reported separated periodic orbit—the 1-point periodic orbit—in the irrotational case [Miller and Youngkins 1997]. The 3-point periodic orbit was found to be stable although its surrounding quasiperiodic region is smaller compared to that of the 1-point periodic orbit. The presence of the 3-point periodic orbit in the rotational case is most likely a result of an angular momentum-induced bifurcation since it was not observed in the case when $L = 0$.

Also, using the Poincare surface of section, we have singled out one trajectory that may be chaotic, the one which tend to fill certain region of the Poincare section. Its power spectrum is distinctly different from that of periodic and quasiperiodic orbits and revealed a broad spectrum—a characteristic of chaos. Indeed, this single trajectory occupies a large portion of the entire phase-space.

The non-equilibrium behavior the rotational concentric spherical shells model has been analyzed up to first order in the Vlasov-Poisson perturbation. The continuum results were compared with N-body simulations. The improvement of the agreement of the continuum limit with the N-shells simulation with increasing N is apparent.

In the future, it will be interesting to do the following:

*Develop faster implementation of the event-driven algorithm*

We did not use an event-driven algorithm for the shells system because of its slowness. In the future it may still be feasible to implement an event-driven algorithm provided that root finding and event detection routines are speeded up. It would be interesting to compare the performance of this still to be implemented event driven algorithm with the time-stepping numerical integrator we introduced in this thesis especially for studying relaxation and other aspects associated with the long time behavior of the system. A heap algorithm by a group from Sweden [47] has been used for the parallel sheets system. The algorithm is a general one used in tracing the dynamics of one-dimensional systems and could therefore be used for the shells system. One problem that has to be addressed is the accumulated truncation errors during calculations of crossing radius. Also, the existing event-driven algorithms are not designed to track multi-particle crossings beyond pair encounters. Future simulations studies could find great help from developing such algorithm that could achieve these requirements.

*Model evaporation process in globular clusters using rotational spheres systems*

Another problem worth mentioning is related to mass and energy loss in globular clusters. The first aspect could be modeled using the rotational shells model. The evaporation process has been modeled using parallel sheets [48], and irrotational spherical shells [7]. In these models an artificial outer boundary is introduced and any sheet/shell hitting the boundary is considered lost to the system. In addition an artificial inner barrier is introduced to shield the singularity in the irrotational spheres system. A mass evaporation model incorporating rotational effects may obviate the need for an artificial barriers or boundaries.

*Study clustering, mass-segregation, and structure formation in the context of spherical systems*

The clustering problem in classical gravitational systems is also a rich problem for research. In fact, various approaches have been used to characterize different types of clustering phenomena in gravitating systems. Some groups have used the straightforward sticking process to study cluster size evolution in one-dimensional systems [49,50]. Mass segregation due to equipartition of energy was also shown as a possible mechanism in clustering [26,12]. Clustering in higher dimensions was studied recently by an Italian group [51]. A systematic study of clustering in spherically symmetric systems may be a useful complement to the abovementioned studies.

# APPENDIX A
# Summary of numerical integrators [52]

## A.1 Euler method

$$a_n = a(r_n)$$
$$r_{n+1} = r_n + v_n \delta t$$
$$v_{n+1} = v_n + a_n \delta t$$

## A.2 Euler-Cromer method

$$a_n = a(r_n)$$
$$v_{n+1} = v_n + a_n \delta t$$
$$r_{n+1} = r_n + v_{n+1} \delta t$$

## A.3 Modified Euler-Cromer method

$$a_n = a(r_n)$$
$$v_{n+1} = v_n + a_n \delta t$$
$$r_{n+1} = r_n + v_{n+1} \delta t + \frac{1}{2} a_n \delta t^2$$

## A.4 Verlet method

$$a_n = a(r_n)$$
$$r_{n+1} = r_n + v_{n+1} \delta t + \frac{1}{2} a_n \delta t^2$$
$$a_{n+1} = a(r_{n+1})$$
$$v_{n+1} = v_n + 0.5(a_n + a_{n+1}) \delta t$$

## A.5 Fourth-order Runge-Kutta method

$$k_{r1} = v_n,\ k_{v1} = a(r_n)$$
$$k_{r2} = v_n + \frac{1}{2} k_{v1} \delta t,\ k_{v2} = a\left(r_n + \frac{1}{2} k_{r1} \delta t\right)$$
$$k_{r3} = v_n + \frac{1}{2} k_{v2} \delta t,\ k_{v3} = a\left(r_n + \frac{1}{2} k_{r2} \delta t\right)$$
$$k_{r4} = v_n + k_{v3} \delta t,\ k_{v4} = a(r_n + k_{r3} \delta t)$$
$$r_{n+1} = r_n + \frac{1}{6}\left(k_{r1} + 2(k_{r2} + k_{r3}) + k_{r4}\right) \delta t,\ v_{n+1} = v_n + \frac{1}{6}\left(k_{v1} + 2(k_{v2} + k_{v3}) + k_{v4}\right) \delta t$$

# APPENDIX B
# Detailed calculations used in Chapter 3

## B.1 Initial phase-space density

$$f(r,v,t=0)=n(r)\varphi(v)$$

$$n(r)=\lim_{b\to\infty}\sqrt{\frac{b}{\pi}}\left\{\exp\left[-b(r-z)^2\right]+\exp\left[-b(r+z)^2\right]\right\}$$

$$\varphi(v)=\lim_{b\to\infty}\frac{1}{4s}\left\{\tanh\left[b(v+s)\right]-\tanh\left[b(v-s)\right]\right\}$$

## B.2 Derivation of $f_1(r, v, t)$

$$\int_0^{r-vt+vs}\left[\int_{-\infty}^{\infty}n(x-us)\phi(u)du\right]dx=\sum_{k=0}^{\infty}\frac{(-1)^k}{(k+1)!}(r-vt+vs)^{k+1}\int_{-\infty}^{\infty}n^{(k)}(us)\phi(u)du$$

$$\int_0^t\left\{\frac{Gm}{(r-vt+vs)^2}\int_0^{r-vt+vs}\left[\int_{-\infty}^{\infty}n(x-us)\phi(u)du\right]dx\right\}ds$$

$$=Gm\int_{-\infty}^{\infty}\phi(u)du\sum_{k=0}^{\infty}\frac{(-1)^k}{(k+1)!}\int_0^t n^{(k)}(us)(r-vt+vs)^{k-1}ds$$

$$\int_0^t\left\{\frac{Gm}{(r-vt+vs)^2}\int_0^{r-vt+vs}\left[\int_{-\infty}^{\infty}n(x-us)\phi(u)du\right]dx\right\}sds$$

$$=Gm\int_{-\infty}^{\infty}\phi(u)du\sum_{k=0}^{\infty}\frac{(-1)^k}{(k+1)!}\int_0^t n^{(k)}(us)(r-vt+vs)^{k-1}sds$$

$$\int_0^t n^{(k)}(us)(r-vt+vs)^{k-1}ds=\frac{1}{u}\int_0^{ut}n^{(k)}(x)\left(r-vt+\frac{v}{u}x\right)^{k-1}dx$$

$$=\frac{1}{u}\sum_{j=1}^{k}\left(-\frac{v}{u}\right)^{k-j}\frac{(k-1)!}{(j-1)!}\left[n^{(j-1)}(ut)r^{j-1}-n^{(j-1)}(0)(r-vt)^{j-1}\right]$$

$$\int_0^t n^{(k)}(us)(r-vt+vs)^{k-1}sds=\frac{1}{u^2}\int_0^{ut}n^{(k)}(x)\left(r-vt+\frac{v}{u}x\right)^{k-1}xdx$$

$$=\frac{1}{u^2}\sum_{j=1}^{k}\left(-\frac{v}{u}\right)^{k-j}\frac{(k-1)!}{(j-1)!}\left\{\begin{bmatrix}utn^{(j-1)}(ut)\\-(k-j+1)n^{(j-2)}(ut)\end{bmatrix}r^{j-1}\atop +(k-j+1)n^{(j-2)}(0)(r-vt)^{j-1}\right\}$$

## B.3 General expressions of $A_k$ and $B_k$

$$A_0=\int_0^{ut}\frac{n(x)}{r-vt+\frac{v}{u}x}dx$$

$$A_1 = n(ut) - n(0)$$

$$A_2 = \left(-\frac{v}{u}\right) A_1 + \left[ n^{(1)}(ut) r - n^{(1)}(0)(r - vt) \right]$$

$$A_3 = 2\left(-\frac{v}{u}\right) A_2 + \left[ n^{(2)}(ut) r^2 - n^{(2)}(0)(r - vt)^2 \right]$$

$$A_4 = 3\left(-\frac{v}{u}\right) A_3 + \left[ n^{(3)}(ut) r^3 - n^{(3)}(0)(r - vt)^3 \right]$$

$$\vdots$$

$$A_k = -(k-1)\left(\frac{v}{u}\right) A_{k-1} + \left[ n^{(k-1)}(ut) r^{k-1} - n^{(k-1)}(0)(r - vt)^{k-1} \right]$$

$$B_0 = \int_0^{ut} \frac{n(x)}{r - vt + \frac{v}{u}x} x\,dx$$

$$B_1 = \left[ utn(ut) - n^{(-1)}(ut) \right] + n^{(-1)}(0)$$

$$B_2 = \left(-\frac{v}{u}\right)\left[ 2B_1 - utn(ut) \right] + \left\{ \left[ utn^{(1)}(ut) - n(ut) \right] r + n(0)(r - vt) \right\}$$

$$B_3 = 2\left(-\frac{v}{u}\right)^2 (-B_1) + 2\left(-\frac{v}{u}\right)\left[ 2B_2 - utn^{(1)}(ut) r \right] + \left\{ \left[ utn^{(2)}(ut) - n^{(1)}(ut) \right] r^2 + n^{(1)}(0)(r - vt)^2 \right\}$$

$$B_4 = 6\left(-\frac{v}{u}\right)^2 (-B_2) + 3\left(-\frac{v}{u}\right)\left[ 2B_3 - utn^{(2)}(ut) r^2 \right] + \left\{ \left[ utn^{(3)}(ut) - n^{(2)}(ut) \right] r^3 + n^{(2)}(0)(r - vt)^3 \right\}$$

$$B_5 = 12\left(-\frac{v}{u}\right)^2 (-B_3) + 4\left(-\frac{v}{u}\right)\left[ 2B_4 - utn^{(3)}(ut) r^3 \right] + \left\{ \left[ utn^{(4)}(ut) - n^{(3)}(ut) \right] r^4 + n^{(3)}(0)(r - vt)^4 \right\}$$

$$\vdots$$

$$B_k = \begin{cases} (k-1)(k-2)\left(\frac{v}{u}\right)^2 B_{k-2} - (k-1)\left(\frac{v}{u}\right)\left[ 2B_{k-1} - utn^{(k-2)}(ut) r^{k-2} \right] \\ + \left\{ \left[ utn^{(k-1)}(ut) - n^{(k-2)}(ut) \right] r^{k-1} + n^{(k-2)}(0)(r - vt)^{k-1} \right\} \end{cases}$$

## B.4 Expressions of $A_k$ and $B_k$ for the expanding spherical shells

$$A_0 = \int_0^{ut} \frac{\delta(x-z) + \delta(x+z)}{r - vt + \frac{v}{u}x} dx = \int_0^{ut} \frac{\delta(x-z)}{r - vt + \frac{v}{u}x} dx - \int_0^{-ut} \frac{\delta(x-z)}{r - vt - \frac{v}{u}x} dx = \frac{\operatorname{sgn}(u)}{r - vt + \frac{v}{|u|}z} H(|u|t - z)$$

$$A_1 = n(ut)$$

$$A_2 = \left(-\frac{v}{u}\right) A_1 + n^{(1)}(ut) r$$

$$A_3 = 2\left(-\frac{v}{u}\right)A_2 + n^{(2)}(ut)r^2$$

$$A_4 = 3\left(-\frac{v}{u}\right)A_3 + n^{(3)}(ut)r^3$$

$$\vdots$$

$$A_k = -(k-1)\left(\frac{v}{u}\right)A_{k-1} + n^{(k-1)}(ut)r^{k-1}$$

$$B_0 = \int_0^{ut} \frac{\delta(x-z)+\delta(x+z)}{r - vt + \frac{v}{u}x} x dx = \int_0^{ut} \frac{\delta(x-z)}{r - vt + \frac{v}{u}x} x dx + \int_0^{-ut} \frac{\delta(x-z)}{r - vt - \frac{v}{u}x} x dx = \frac{z}{r - vt + \frac{v}{|u|}z} H(|u|t - z)$$

$$B_1 = utn(ut) - n^{(-1)}(ut) = utn(ut) - \left[H(ut+z) + H(ut-z) - 1\right]$$

$$B_2 = \left(-\frac{v}{u}\right)\left[2B_1 - utn(ut)\right] + \left[utn^{(1)}(ut) - n(ut)\right]r$$

$$B_3 = 2\left(\frac{v}{u}\right)^2(-B_1) - 2\left(\frac{v}{u}\right)\left[2B_2 - utn^{(1)}(ut)r\right] + \left[utn^{(2)}(ut) - n^{(1)}(ut)\right]r^2$$

$$B_4 = 6\left(\frac{v}{u}\right)^2(-B_2) - 3\left(\frac{v}{u}\right)\left[2B_3 - utn^{(2)}(ut)r^2\right] + \left[utn^{(3)}(ut) - n^{(2)}(ut)\right]r^3$$

$$\vdots$$

$$B_k = (k-1)(k-2)\left(\frac{v}{u}\right)^2 B_{k-2} - (k-1)\left(\frac{v}{u}\right)\left[2B_{k-1} - utn^{(k-2)}(ut)r^{k-2}\right] + \left[utn^{(k-1)}(ut) - n^{(k-2)}(ut)\right]r^{k-1}$$

## B.5 First five derivatives of *n*(*r*)

$$n^{(1)}(r) = -2b\left\{\sqrt{\frac{b}{\pi}}\left[(r-z)\exp\left[-b(r-z)^2\right] + (r+z)\exp\left[-b(r+z)^2\right]\right]\right\}$$

$$n^{(2)}(r) = -2b\begin{Bmatrix} n(r) \\ -2b\sqrt{\frac{b}{\pi}}\left[(r-z)^2\exp\left[-b(r-z)^2\right] + (r+z)^2\exp\left[-b(r+z)^2\right]\right] \end{Bmatrix}$$

$$n^{(3)}(r) = -2b\begin{Bmatrix} 3n^{(1)}(r) \\ +4b^2\sqrt{\frac{b}{\pi}}\left[(r-z)^3\exp\left[-b(r-z)^2\right] + (r+z)^3\exp\left[-b(r+z)^2\right]\right] \end{Bmatrix}$$

$$n^{(4)}(r) = -2b\begin{Bmatrix} 6n^{(2)}(r) \\ -8b^3\sqrt{\frac{b}{\pi}}\left[(r-z)^4\exp\left[-b(r-z)^2\right] + (r+z)^4\exp\left[-b(r+z)^2\right]\right] \end{Bmatrix}$$

$$n^{(5)}(r) = -2b\begin{Bmatrix} 10n^{(3)}(r) + 30bn^{(1)}(r) \\ +16b^4\sqrt{\frac{b}{\pi}}\left[(r-z)^5\exp\left[-b(r-z)^2\right] + (r+z)^5\exp\left[-b(r+z)^2\right]\right] \end{Bmatrix}$$

# APPENDIX C
# Source codes

## C.1 Numerical simulation of 2-shell system

```
%%%%%%%%%% START OF MATLAB CODE %%%%%%%%%%
% Lines that begin with "%" sign are remarks and non-executable
clear
format long g
% INITIALIZATION OF PARAMETERS
Nshell = 2;  % Number of particles
G = 1; % Gravitational constant
L = 1; % Angular momentum magnitude of stars in a shell
m = 1; % Effective mass of a shell
E0min = -G^2*m^5/L^2;  % Least possible total energy
Emin = 9*E0min/8;   % Least possible energy of a shell
E0max = Emin/9;  % Maximum possible total energy of a bounded system
dt = 0.001;  % Time step size
tmax = 5000;  % Maximum dynamical time
MaxColl = 50000; % Maximum number of collisions/crossings
E0 = -0.250; % Total energy of the system
R1 = (-G*m^2/E0)*0.50001; % Initial crossing radius
r = R1*[1 1];% Initial radius
v = -sqrt((E0 + (3*G*m^2)^2/(8*Emin*R1^2) + 2*G*m^2/R1)/m)*[1 -1]; % Initial velocity
n = [0 1];   % Initial number of inner shells
g(1) = (L^2/m^2)/r(1)^3 - G*m*(0.5 + n(1))/r(1)^2;   % Corresponding initial
acceleration
g(2) = (L^2/m^2)/r(2)^3 - G*m*(0.5 + n(2))/r(2)^2;
% Note that r, v, n, and g are 2-dimensional column vectors
t = 0; % Initial time
NumColl = 0;  % Current number of collisions/crossings
E1 = 0.5*m*v(1)^2 + L^2/(2*m*r(1)^2) - G*m^2*(0.5 + n(1))/r(1)  % Energy of shell 1
E2 = 0.5*m*v(2)^2 + L^2/(2*m*r(2)^2) - G*m^2*(0.5 + n(2))/r(2)  % Energy of shell 2
E = E1 + E2      % Actual total energy
pause
filename = 'c:\windows\desktop\shell2CHAOTIC5time.txt';
fid = fopen(filename,'w');
starttime = cputime;
% END OF INITIALIZATION
while (t<tmax)
   KE = 0.5*m*v*v';
   E = KE+(L^2/(2*m))*(1/r(1)^2+1/r(2)^2)-G*m^2*((0.5+n(1))/r(1)+(0.5+n(2))/r(2));
   if rem(round(10*t/dt),round(1/dt)) == 0
       fprintf(fid,'%.10f\t%.10f\t%.10f\t%.10f\t%.10f\n',[t; r(1); v(1); r(2);
v(2)]);
       % Record current phase-space coordinates
       % Sampling is every 0.1 time units
   end
   rold = r;
   t = t + dt;
   if E < E0
   %%%%%%%%% VERLET METHOD %%%%%%%%%%
   r = r + v*dt + 0.5*g*dt*dt;
   gnew(1) = (L^2/m^2)/r(1)^3 - G*m*(0.5 + n(1))/r(1)^2;
   gnew(2) = (L^2/m^2)/r(2)^3 - G*m*(0.5 + n(2))/r(2)^2;
   v = v + 0.5*(g+gnew)*dt;
   g = gnew;
   %%%%%%%% END VERLET METHOD %%%%%%%%%
   else
   %%%%%%%% MODIFIED EULER-CROMER METHOD %%%%%%%%%%
   v = v + g*dt;
   r = r + v*dt + 0.5*g*dt*dt;
   g(1) = (L^2/m^2)/r(1)^3 - G*m*(1/2 + n(1))/r(1)^2;
   g(2) = (L^2/m^2)/r(2)^3 - G*m*(1/2 + n(2))/r(2)^2;
   %%%%%%%% END MODIFIED EULER-CROMER METHOD %%%%%%%%%%
```

```
    end
    %%%%%%%%%%% SIMPLE EULER METHOD %%%%%%%%%%%
    % r = r + v*dt;
    % v = v + g*dt;
    % g(1) = (L^2/m^2)/r(1)^3 - G*m*(1/2 + n(1))/r(1)^2;
    % g(2) = (L^2/m^2)/r(2)^3 - G*m*(1/2 + n(2))/r(2)^2;
    %%%%%%%%%%% END SIMPLE EULER METHOD %%%%%%%%%%%
    %%%%%%%%%%% EULER-CROMER METHOD %%%%%%%%%%%
    % v = v + g*dt;
    % r = r + v*dt;
    % g(1) = (L^2/m^2)/r(1)^3 - G*m*(1/2 + n(1))/r(1)^2;
    % g(2) = (L^2/m^2)/r(2)^3 - G*m*(1/2 + n(2))/r(2)^2;
    %%%%%%%%%%% END EULER-CROMER METHOD %%%%%%%%%%%
    %%%%%%%%%%% FOURTH RUNGE-KUTTA %%%%%%%%%%%
    % g(1) = (L^2/m^2)/r(1)^3 - G*m*(1/2 + n(1))/r(1)^2;
    % g(2) = (L^2/m^2)/r(2)^3 - G*m*(1/2 + n(2))/r(2)^2;
    % k1v = g*dt;
    % k1r = v*dt;
    % g(1) = (L^2/m^2)/(r(1)+k1r(1)/2)^3 - G*m*(1/2 + n(1))/(r(1)+k1r(1)/2)^2;
    % g(2) = (L^2/m^2)/(r(2)+k1r(2)/2)^3 - G*m*(1/2 + n(2))/(r(2)+k1r(2)/2)^2;
    % k2v = g*dt;
    % k2r = (v+k1v/2)*dt;
    % g(1) = (L^2/m^2)/(r(1)+k2r(1)/2)^3 - G*m*(1/2 + n(1))/(r(1)+k2r(1)/2)^2;
    % g(2) = (L^2/m^2)/(r(2)+k2r(2)/2)^3 - G*m*(1/2 + n(2))/(r(2)+k2r(2)/2)^2;
    % k3v = g*dt;
    % k3r = (v+k2v/2)*dt;
    % g(1) = (L^2/m^2)/(r(1)+k3r(1))^3 - G*m*(1/2 + n(1))/(r(1)+k3r(1))^2;
    % g(2) = (L^2/m^2)/(r(2)+k3r(2))^3 - G*m*(1/2 + n(2))/(r(2)+k3r(2))^2;
    % k4v = g*dt;
    % k4r = (v+k3v)*dt;
    % v = v + (k1v + 2*k2v + 2*k3v + k4v)/6;
    % r = r + (k1r + 2*k2r + 2*k3r + k4r)/6;
    %%%%%%%%%%% END FOURTH RUNGE KUTTA %%%%%%%%%%%
    if (rold(1) - rold(2))*(r(1) - r(2)) < 0
        % Check for occurrence of collision/crossing
        n(1) = n(2);
        n(2) = 1-n(1);
        gbuff = g(1);
        g(1) = g(2);
        g(2) = gbuff;
        NumColl = NumColl + 1
    end
end
runningtime = cputime – starttime % Displays approximate running time in seconds
fclose(fid);
%%%%%%%%%%% END OF MATLAB CODE %%%%%%%%%%%
```

## C.2 Numerical simulation of *N*-shell system

```
%%%%%%%%%%% START OF MATLAB CODE %%%%%%%%%%%
% Lines that begin with "%" sign are remarks and non-executable
clear
format long g
% INITIALIZATION OF PARAMETERS
N = 1024; % Number of shells
G = 1; % Gravitational constant
Lt = 1;   % Constant related to angular momentum
M = 1; % Total mass of the system
L = Lt/N; % Angular momentum magnitude of every shell
m = M/N;  % Mass of every shell
tmax = 500;      % Maximum dynamical time
E = 1e-0  % Total energy of the system
l = 0.05  % Reciprocal of the initial virial ratio
P = 2*l*E/(2*l - 1);   % Initial potential energy
z = -G*M^2/(2*P) % Radius parameter
s = sqrt((3*G*M/(2*z))*((N-1)/(N+1))*(1/l - 2*Lt^2/(G*M^3*z)))  % Velocity parameter
T = z/s      % Time constant
```

```
dt = 0.001;
pause;
t = 0; % Initial time
r = z*linspace(1,N,N)./linspace(1,N,N); % Initial radius
v = linspace(-s,s,N);   % Initial velocity
n = linspace(0,N-1,N); % Initial number of inner shells
a = (L^2/m^2)./r.^3 - G*m*(1/2 + n)./r.^2; % Initial acceleration
aold = a; % Acceleration buffer
filename = 'c:\windows\desktop\data.txt';
fid = fopen(filename,'w');
fprintf(fid, '%.10f\t%.10f\n', [r; v]);
hold on;
plot(r,v,'-');   % Plot initial phase-space coordinates
starttime = cputime;
for c=0.1:0.1:1
    c   % Display sampling time
    while t < c*T
       % Find phase-space coordinates at time t+dt
       t = t + dt;
       r = r + v*dt + 0.5*a*dt^2;
       v = v + a*dt + 0.5*(a - aold)*dt;
       aold = a;
       a = (L^2/m^2)./r.^3 - G*m*(1/2 + n)./r.^2;
   end
   fprintf(fid, '%.10f\t%.10f\n', [r; v]);
end
runningtime = (cputime - starttime)/60  % Display total running time in seconds
plot(r,v,'-o');  % Plot final phase-space coordinates
fclose(fid);
%%%%%%%%%% END OF MATLAB CODE %%%%%%%%%%
```